\documentclass[aps,pra,twocolumn,superscriptaddress,nofootinbib]{revtex4-2}

\usepackage{amsmath,amsthm,amsfonts,amssymb,bm,graphicx,color,mathpazo,times, braket}
\usepackage[colorlinks={true}, citecolor={blue}, filecolor={blue}, linkcolor={blue}, urlcolor={blue}]{hyperref}
\usepackage[caption=false]{subfig}
\usepackage{xcolor}
\usepackage{float}
\usepackage{placeins}
\usepackage{soul}
\usepackage{url}

\begin{document}

\title{Topological Engine Monitor: Persistent Homology-Based Fault Detection in Finite-Time Quantum Engines}

\author{Mira\c{c} Kerem Maden}
\email[Corresponding author: ]{mmaden25@ku.edu.tr}
\affiliation{Department of Physics, Ko\c{c} University, 34450 Sar{\i}yer, Istanbul, T\"{u}rkiye}

\author{Asghar Ullah}
\affiliation{Department of Physics, Ko\c{c} University, 34450 Sar{\i}yer, Istanbul, T\"{u}rkiye}

\author{Baris Coskunuzer}
\affiliation{Department of Mathematical Sciences, The University of Texas at Dallas, Richardson, TX 75080, USA}

\author{\"Ozg\"ur E. M\"ustecapl\i o\u{g}lu}
\affiliation{Department of Physics, Ko\c{c} University, 34450 Sar{\i}yer, Istanbul, T\"{u}rkiye}
\affiliation{T\"UBITAK Research Institute for Fundamental Sciences (TBAE), 41470 Gebze, T\"{u}rkiye}

\begin{abstract}

The reliable operation of finite-time quantum heat engines is fundamentally limited by control imperfections that induce nonadiabatic phase accumulation and quantum friction, degrading the stability of the thermodynamic cycle. Traditional monitoring relies on energetic observables such as instantaneous cycle work; however, under finite-time driving, these quantities exhibit strong fluctuations, obscuring reliable short-window detection of control degradation without extensive statistical averaging. Here, we apply a topological data analysis (TDA)-based approach to establish a measurement-efficient, geometric framework for diagnosing control degradation in finite-time quantum Otto engines. We construct time-delay embeddings from an idealized continuous record of a single observable and map the reconstructed dynamics into persistent homology diagrams. We define a scalar quality index based on Wasserstein and Bottleneck distances that tracks control degradation and anticipates the loss of stable cyclic operation. By encoding topology via persistence images and silhouettes, we achieve highly robust classification of degraded operation across diverse noise profiles. We benchmark the TDA-based approach (topological engine monitor, TEM) against a standard multi-feature statistical baseline (spectral-statistical monitor, SSM) across progressively structured and localized noise settings, from global timing jitter to correlated adiabatic noise and coherence injection. We find that as the perturbations become more structured and localized, the conventional SSM approach degrades while the TEM remains robust. Finally, a pixel-wise Pearson correlation analysis reveals that the method captures microscopic signatures of quantum friction. Our results demonstrate the potential of topology-based diagnostics for non-ideal quantum thermodynamic devices.
\end{abstract}

\maketitle

\section{INTRODUCTION}
\label{sec:intro}

Quantum thermodynamics seeks to extend the classical laws of heat and work into the regime where quantum fluctuations, entanglement, and coherence dominate system dynamics \cite{kosloff2013, alicki1979, Vinjanampathy01102016,Goold_2016}. Over the past decade, significant progress has been made in realizing continuous quantum heat engines, autonomous refrigerators, and periodically driven thermal machines \cite{PhysRevE.76.031105, PhysRevB.96.104304,PhysRevLett.127.200602,PhysRevLett.123.240601, PhysRevLett.2.262,PhysRevLett.129.100603,Johannes2016,PhysRevLett.122.110601, uzdin2015}. A central challenge is that a reversible operation necessarily implies vanishing power output. Consequently, to extract macroscopic, non-vanishing power from these microscopic systems, they must operate in a strictly finite-time regime \cite{PhysRevE.68.016101,kosloff2013,Dann2020}. For generic finite-time protocols involving noncommuting Hamiltonians, rapid driving can induce transitions between instantaneous energy eigenstates and generate quantum friction~\cite{kato1950,messiah2014quantum,PhysRevLett.113.260601,PhysRevLett.125.180603}. The resulting coherences in the energy basis may dissipate as heat during the subsequent thermalization strokes. This friction fundamentally degrades the engine performance by introducing irreversibility during the unitary strokes, reducing both efficiency and power output well below the Carnot and Curzon-Ahlborn bounds~\cite{Rezek_2006, PhysRevLett.105.150603,PhysRevLett.109.203006}. Therefore, as quantum technologies transition from proof-of-concept models to scalable, autonomous devices, the ability to monitor their operational health and rapidly detect control degradation arising from timing jitter, drive-amplitude noise, or environmental decoherence becomes increasingly important. This challenge is particularly acute in quantum heat engines and refrigerators, where microscopic control errors can propagate across thermodynamic cycles. In classical machines, faults are typically detected by monitoring real-time power output and identifying drops below a predefined threshold. Unfortunately, applying this standard energy-tracking approach to quantum engines is fundamentally unreliable. Because thermalization with the reservoirs is incomplete and unitary strokes are highly nonadiabatic, single-cycle observables—such as discrete cycle work and heat exchange—are intrinsically stochastic~\cite{RevModPhys.83.771, RevModPhys.81.1665, Vinjanampathy01102016}. They exhibit massive cycle-to-cycle variance, precluding the real-time detection of anomalies. Relying on mean energy output requires extensive temporal averaging, which defeats the purpose of rapid, real-time diagnostics. Consequently, alternative diagnostics that analyze the structure of fluctuations rather than their mean values are required for reliable real-time monitoring of quantum engines.

To overcome the limitations of thermodynamic measurement, we propose a paradigm shift: decoupling condition monitoring from energy tracking and instead monitoring the global geometric structure of the engine's phase space. By treating an idealized continuous record of a single quantum observable as a complex dynamical system, we can reconstruct the underlying operational manifold using time-delay embeddings \cite{takens1981}. The degradation of the engine's limit cycle—driven by quantum friction and timing jitter—manifests not only as energetic noise but also as structural deformation. To quantify this structural breakdown, we apply topological data analysis (TDA), in particular, persistent homology~\cite{edelsbrunner2008, carlsson2009}. Persistent homology has recently emerged as a promising tool for the analysis of quantum dynamical systems, including quantum chaos \cite{Cao2023}, nonequilibrium many-body dynamics \cite{Spitz2021, Leykam2023, Spitz2025}, and topology-aware quantum-data representations \cite{gvys-hl8h, Scali2024}. TDA is highly robust against local continuous deformations, making it uniquely suited for extracting invariant large-scale features from irregular quantum trajectories. While TDA has found profound success in condensed matter physics for identifying topological phases of matter \cite{macpherson2020}, its application as a diagnostic tool for control quality for non-equilibrium quantum thermodynamics remains largely unexplored.

In this work, we apply a TDA-based approach to monitor the performance of a finite-time quantum Otto engine under control degradation. We first demonstrate that traditional energetic diagnostics (e.g., work) exhibit large fluctuations that mask engine degradation. To mitigate this, we introduce a scalar quality index (QI), based on Wasserstein and Bottleneck distances, that tracks the breakdown of the engine’s limit cycle. We further encode topology via persistence images and silhouettes, capturing the progressive smearing of the operational manifold.
We benchmark this TDA-based approach, hereafter referred to as the topological engine monitor (TEM), against a standard multi-feature statistical baseline, referred to as the spectral-statistical monitor (SSM), across five progressively structured and localized perturbation settings, ranging from global cycle corruption to localized, correlated adiabatic noise. As noise becomes more structured and bandwidth-limited, the SSM degrades while the TEM retains strong discriminative power. Utilizing these representations, we construct a machine-learning pipeline within the TEM framework that achieves robust discrimination between nominal and degraded operation. Finally, we perform a pixel-wise Pearson correlation analysis on the persistence images, indicating that quantum friction manifests as high-frequency micro-loops rather than uniform phase-space expansion. We emphasize that the present work focuses on condition monitoring rather than the suppression of nonadiabatic dynamics. The nominal cycle considered here is itself a finite-time, generally nonadiabatic cycle, whereas the perturbations studied are additional deviations of the implemented controls from their prescribed values, arising from timing jitter, waveform distortion, finite bandwidth, calibration drift, or residual coupling. These perturbations are introduced only to generate labeled benchmark data, while TEM identifies degradation solely from the measured engine trajectories. Counterdiabatic driving and related shortcut-to-adiabaticity protocols provide alternative strategies for reducing unwanted finite-time excitations and quantum friction~\cite{berry2009transitionless,torrontegui2013shortcuts,delcampo2014superadiabatic,cakmak2019spin,pedram2023quantum}. Such methods modify the control protocol, whereas TEM diagnoses additional departures of the measured trajectory from the nominal finite-time reference dynamics. We stress that no counterdiabatic or shortcut-to-adiabaticity protocol is implemented or evaluated in the present study.

The rest of the paper is organized as follows. Section~\ref{sec:system} introduces the model and numerical methods, including the open-system dynamics of the qubit working medium, the finite-time quantum Otto cycle, the control-degradation models, and the numerical integration procedure. Section~\ref{sec:tda} presents the complete TDA and machine-learning framework. Section~\ref{sec:results} presents and discusses the results under the different noise settings, including the comparison between the SSM and TEM frameworks. Finally, Sec.~\ref{sec:conclusion} summarizes our conclusions and outlook. We discuss the SSM in Appendix~\ref{app:ssm} and provide supporting technical details in Appendices~\ref{app:tda_params} and~\ref{app:silhouettes}.

\section{Model and Methods}
\label{sec:system}

\subsection{Qubit working medium and open system dynamics}
The working medium of the engine is a single two-level system (qubit) governed by the time-dependent Hamiltonian (we set $k_B = \hbar = 1$ throughout this work):
\begin{equation}
    H(t) = \frac{1}{2}\left[\omega_z(t)\sigma_z + \omega_x(t)\sigma_x\right],
    \label{eq:hamiltonian}
\end{equation}
where $\sigma_x$ and $\sigma_z$ are the standard Pauli matrices. The transverse field $\omega_x(t)$ acts as a controllable source of quantum coherence. Because $[H(t),H(t')] \neq 0$ at different times during the drive, the dynamics can generate nonadiabatic transitions. In an ideal quantum Otto engine, the unitary strokes are implemented adiabatically by slowly varying $\omega_z(t)$, thereby suppressing such transitions between instantaneous energy eigenstates. In realistic settings, however, finite ramp speeds, residual couplings, or control imperfections inevitably introduce effective noncommuting contributions. The transverse term $\omega_x(t)\sigma_x$ therefore provides a minimal tunable model capturing quantum friction and coherence generation during finite-time operation.

The state of the system is parameterized by its Bloch vector $\mathbf{r}(t) = (x(t), y(t), z(t))^T$, where $x_i(t) = \langle\sigma_i\rangle = \mathrm{Tr}[\rho(t)\sigma_i]$. 

During isolated, coherent evolution (the work strokes), the system obeys the von Neumann equation, which yields the unitary Bloch equations:
\begin{equation}
\label{eq:bloch_unitary}
\begin{aligned}
    \dot{x}(t) &= -\omega_z(t) y(t), \\
    \dot{y}(t) &= \omega_z(t) x(t) - \omega_x(t) z(t), \\
    \dot{z}(t) &= \omega_x(t) y(t).
\end{aligned}
\end{equation}
These equations show that when both $\omega_x(t)\neq0$ and $\omega_z(t)\neq0$, the noncommuting Hamiltonian generates a finite $y$ component, corresponding to the creation of quantum coherence in the instantaneous energy eigenbasis during finite-time driving.

During the isochoric strokes, the qubit is weakly coupled to a thermal reservoir at temperature $T$. The dissipative dynamics are governed by a Markovian master equation in the Lindblad form. In the Bloch representation, this yields a relaxation toward the instantaneous Gibbs state at a characteristic rate $\Gamma$:
\begin{equation}
\label{eq:bloch_ME}
\begin{aligned}
    \dot{x}(t) &= -\frac{\Gamma}{2}x(t) - \omega_z(t)y(t), \\
    \dot{y}(t) &= \omega_z(t)x(t) - \frac{\Gamma}{2}y(t), \\
    \dot{z}(t) &= -\Gamma\left[z(t) - z_{eq}(\omega_z, T)\right],
\end{aligned}
\end{equation}
where the thermal fixed point is
\begin{equation}
    z_{eq}(\omega_z, T) = -\tanh(\omega_z / 2T).
\end{equation}
Note that the transverse relaxation rates (dephasing) are dictated by $\Gamma_2 = \Gamma/2$, while the longitudinal relaxation follows $\Gamma_1 = \Gamma$. The above Bloch-equation description constitutes an effective weak-coupling, Markovian approximation consistent with a Lindblad master equation formulated in the energy eigenbasis of the isochoric Hamiltonian \cite{breuer2002theory}.

\subsection{The four-stroke cycle and timing jitter}
The engine operates according to a four-stroke finite-time
quantum Otto cycle, implemented through explicit time-dependent control of the qubit Hamiltonian parameters. Each cycle consists of the following stages:

\textit{Hot isochore:} 
The qubit is coupled to a hot reservoir at temperature $T_h$ for a duration $\tau_h$ with a fixed longitudinal field $\omega_z = \omega_h$ and $\omega_x = 0$. During this stage, the system relaxes toward the Gibbs state of $H_h=\omega_h\sigma_z/2$. Since $\tau_h \sim \Gamma^{-1}$, thermalization is incomplete, allowing memory of previous strokes to persist.

\textit{Expansion stroke:} 
The qubit is isolated from the reservoir and evolves unitarily for a duration $\tau_1$. The parameters are smoothly varied according to a rescaled time $s = t/\tau_1 \in [0,1]$:
\begin{align}
\omega_z(s) &= \omega_h + (\omega_c - \omega_h)s, \\
\omega_x(s) &= \omega_x^{\max} \sin(\pi s).
\end{align}
The transverse control $\omega_x(s)$ is activated only during the ramp, which ensures that the Hamiltonian does not commute at different times. The resulting coherences constitute internal quantum friction and play a central role in determining the engine dynamics.

\textit{Cold isochore:} The qubit is coupled to a cold reservoir at temperature $T_c$ for a finite duration $\tau_c$ which implies that thermalization may again be incomplete with $\omega_z = \omega_c$ and $\omega_x = 0$.

\textit{Compression stroke:}
The qubit is isolated from the reservoir and driven in reverse from $\omega_c$ back to $\omega_h$ over a duration $\tau_3$, employing the same sinusoidal envelope for $\omega_x(s)$.

The durations $\tau_h$, $\tau_1$, $\tau_c$, and $\tau_3$ determine the characteristic time scales of the cycle. In the regime considered here, the unitary strokes are neither adiabatic nor sudden, while the isochoric strokes do not lead to complete thermalization. As a result, coherence generation, incomplete relaxation, and memory effects coexist and jointly influence engine performance. To model realistic control imperfections, we introduce independent Gaussian timing jitter into the stroke durations. The nominal duration $\tau_j$ is perturbed as 

\begin{equation}   
\tau_j \rightarrow \tau_j(1 + \delta_j), \quad \delta_j \sim \mathcal{N}(0, \sigma_\tau^2).     \label{eq:timing_jitter} \end{equation} 

This stochasticity breaks the strict periodicity of the driving protocol, preventing the system from settling into a perfect limit cycle.

The timing-jitter parameter $\sigma_\tau$ is varied continuously over the interval $[0,0.25]$, corresponding to relative timing fluctuations ranging from $0\%$ to $25\%$. This interval was selected as a stress-test range spanning nominal to strongly degraded operation. The classification threshold $\sigma_{\mathrm{th}}=0.125$ is the midpoint of this interval and produces approximately balanced classes; it is not interpreted as a universal physical critical value. Thresholds of $\sigma_{\mathrm{th}}=0.10$, $0.125$, and $0.15$ produced persistence-image AUC values spanning $0.8418$--$0.9176$. We retain the midpoint threshold to avoid class imbalance and post-hoc threshold optimization.

\subsection{Models of control degradation}
\label{sec:degradation_models}

In addition to Gaussian timing jitter in the total stroke durations, we consider control perturbations that directly deform the geometric structure of the engine trajectory in Bloch space. These perturbations modify the local nonadiabatic phase accumulation and therefore alter the topology of the reconstructed phase-space manifold. To benchmark our TEM pipeline, we construct the following localized, physically motivated benchmark perturbations.

The noise variables introduced below are used only to generate controlled benchmark trajectories and ground-truth labels. They are not provided as inputs to TEM, which receives only the measured scalar signal $x(t)=\langle\sigma_x(t)\rangle$ and determines whether the realized dynamics have departed from the nominal operating regime. Accordingly, the labels ``nominal'' and ``degraded'' distinguish the absence or presence of additional control perturbations; they do not distinguish adiabatic from nonadiabatic engine operation.

\textit{1. Adiabatic ramp distortion:} 
\label{sec:ramp_distortion_theory}
Instead of a linear interpolation, we introduce stochastic variations in the sweep profile of the longitudinal field. The control during the expansion stroke is parameterized as
\begin{equation}
    \omega_z(s) = \omega_h + (\omega_c - \omega_h) s^{\alpha_n}, \quad s \in [0,1],
    \label{eq:ramp_omega}
\end{equation}
where the linear progression $s$ is distorted by a static exponent $\alpha_n$ drawn per cycle from a normal distribution,
\begin{equation}
    \alpha_n = 1 + \delta_{\alpha,n}, \quad \delta_{\alpha,n} \sim \mathcal{N}(0, \sigma_\alpha^2).
    \label{eq:ramp_alpha}
\end{equation}
This modifies the instantaneous sweep rate as
\begin{equation}
    \dot{\omega}_z(t) \propto \frac{\alpha_n}{\tau_1} s^{\alpha_n-1},
    \label{eq:ramp_deriv}
\end{equation}
thereby directly perturbing the nonadiabatic coherence generation during the unitary strokes.

\textit{2. Correlated adiabatic sweep noise:} 
\label{sec:ou_noise_theory}
In fully controlled engine operations, control-field ramp shapes are rarely subjected to static or completely uncorrelated white noise. To capture finite-memory distortions originating from imperfect waveform synthesis and feedback bandwidth limitations \cite{dechecchi2025dynamics, cantone2025machine, aguilar2008effect}, we employ an Ornstein-Uhlenbeck (OU) stochastic process \cite{stefanatos2020robustness} to perturb the sweep exponent $\alpha(s)$ continuously during the unitary strokes. For instance, the expansion stroke is parameterized as shown in Eq.~\eqref{eq:ramp_omega}, where the local exponent evolves according to the stochastic differential equation
\begin{equation}
    d\alpha(s) = \theta (\mu - \alpha(s)) ds + \sigma_{\mathrm{eff}} dW(s),
    \label{eq:ou_sde}
\end{equation}
During the integration, the exponent is bounded by $\alpha(s)\geq\alpha_{\min}=0.01$. For $\alpha<0$, $s^\alpha$ diverges as $s\rightarrow0$, while $\alpha=0$ reduces the sweep profile to $s^\alpha=1$ away from the initial endpoint. The lower bound therefore serves as a numerical safeguard rather than a physical control parameter and is chosen to be small relative to the nominal value $\alpha=1$. The undistorted linear ramp corresponds to $s^\alpha=s$, and therefore to $\alpha=1$. We consequently set $\mu=1.0$ so that the OU process mean-reverts toward the nominal control profile. The use of an OU process to represent finite-memory control fluctuations follows established stochastic-control descriptions \cite{cantone2025machine,dechecchi2025dynamics,stefanatos2020robustness}. We choose $\theta=50.0$ empirically to represent rapid self-correction, corresponding to the normalized correlation time $\tau_{\mathrm{OU}}=1/\theta=0.02$. Across $\theta=40$, $50$, and $60$, the persistence-image AUC remained between $0.9674$ and $0.9759$, confirming robustness to moderate changes in the mean-reversion rate. For the continuous OU process, the stationary variance scales as $\sigma_{\mathrm{eff}}^2/(2\theta)$ \cite{dechecchi2025dynamics,stefanatos2020robustness}. We therefore set $\sigma_{\mathrm{eff}}=8\sigma_{\mathrm{OU}}$ empirically so that rapid mean reversion does not suppress the imposed fluctuations. For $\sigma_{\mathrm{eff}}=c\sigma_{\mathrm{OU}}$ with $c=6$, $8$, and $10$, the persistence-image AUC remained between $0.9678$ and $0.9713$, confirming robustness to this scale factor. Here, $dW(s)$ represents the standard Wiener process increment. Because the OU process continuously mean-reverts toward the nominal exponent, the perturbation repeatedly corrects itself and largely preserves the global macroscopic boundaries and total duration of the thermodynamic cycle.

\textit{3. Longitudinal high-frequency ripple:} 
\label{sec:ripple_theory}
We also consider a highly specific, phase-coherent perturbation to simulate unintended resonant coupling or harmonic distortion. We inject a high-frequency sinusoidal ripple directly into the longitudinal control field. During the unitary strokes, the field is parameterized as
\begin{equation}
    \omega_z(s) = \omega_{\mathrm{base}}(s) + \delta_{z,n} \sin(k \pi s),
    \label{eq:ripple_omega}
\end{equation}
where $\omega_{\mathrm{base}}(s)$ is the underlying ramp (e.g., linear or distorted), $k$ is an even integer wave number ($k=10$ for expansion and $k=2$ for compression), and the ripple amplitude fluctuates cycle-to-cycle as $\delta_{z,n} \sim \mathcal{N}(0, \sigma_{\mathrm{ripple}}^2)$.

\textit{4. Combined multi-channel control degradation:}
\label{sec:combined_theory}
To construct a stringent multi-channel benchmark, we combine all the preceding control perturbations: global timing jitter, static ramp distortion, finite-bandwidth colored noise, and coherent cross-talk.

\subsection{Thermodynamic definitions and work variance}
\label{sec:thermodynamics}

In the context of explicitly time-dependent Hamiltonians, the instantaneous power delivered by the drive is defined as follows:
\begin{equation}
    \dot{W}(t) = \mathrm{Tr}\left[ \rho(t) \dot{H}(t) \right].
\end{equation}
The work is accumulated discretely over the unitary strokes. The internal energy of the system at step $k$ is given by:
\begin{equation}
    E_k = \mathrm{Tr}[\rho_k H_k] = \frac{1}{2} \left( \omega_z(t_k) z_k + \omega_x(t_k) x_k \right),
\end{equation}
where $\rho_k$ is the system’s density matrix at step $k$, $H_k$ is the corresponding Hamiltonian, and $x_k, z_k$ are the expectation values of the relevant observables.

The differential work performed during a single time step $\Delta t$ is $\delta W_k = E_{k+1} - E_k$, determined under the unitary update before the Hamiltonian parameters are shifted for the next step. The total cycle work $W_n$ for the $n$-th cycle is the sum of the accumulated work across the expansion and compression strokes. 

To quantify the energetic fluctuations induced by timing jitter $\sigma_\tau$, we compute the empirical mean and variance of work over $N$ steady-state cycles:
\begin{equation}
\begin{aligned}
    \bar{W}(\sigma_\tau) &= \frac{1}{N} \sum_{n=1}^{N} W_n, \\
    \mathrm{Var}_W(\sigma_\tau) &= \frac{1}{N} \sum_{n=1}^{N} \left( W_n - \bar{W} \right)^2.
    \end{aligned}
\end{equation}
These quantities quantify the engine performance and illustrate the intrinsic fluctuations arising from finite-time driving and incomplete thermalization.

\subsection{Numerical integration of the master equation}
\label{sec:numerical_integration}

The dynamics of the quantum Otto cycle were simulated by numerically integrating the Bloch equations using a discretized Euler scheme. For a generic driving stroke of duration $\tau_{stroke}$ divided into $N_\text{steps}$ discrete intervals, the time step is $\Delta t = \tau_\text{stroke} / N_\text{steps}$.

We used $N_{\mathrm{steps}}=40$ intervals per stroke, corresponding to $\Delta t=0.015$ for the unitary strokes. Calculations using $N_{\mathrm{steps}}=20$, $40$, and $80$ yielded persistence-image AUC values of $0.8713$, $0.8782$, and $0.8756$, respectively. After resampling onto a common phase grid, the normalized trajectory differences relative to the $40$-step calculation were $0.01840$ and $0.00866$ for $20$ and $80$ steps, respectively. These results confirm that the reported diagnostic performance is insensitive to moderate refinement of the integration grid.

During the isolated unitary strokes (expansion and compression), the dissipative terms vanish ($\Gamma = 0$), and the evolution is purely coherent. The Bloch vector $\mathbf{r}_k$ at time $t_k$ is updated according to:
\begin{equation}
    \mathbf{r}_{k+1} = \mathbf{r}_k + \Delta t \left( \boldsymbol{\omega}_k \times \mathbf{r}_k \right),
\end{equation}
where $\boldsymbol{\omega}_k = (\omega_x(t_k), 0, \omega_z(t_k))^T$ represents the instantaneous drive parameters.

During the isochoric strokes (hot and cold thermalization), the transverse field is turned off ($\omega_x = 0$), and the system undergoes Markovian relaxation. The update rules explicitly decouple the transverse dephasing and longitudinal relaxation as follows:
\begin{align}
    x_{k+1} &= x_k + \Delta t \left( -\frac{\Gamma}{2}x_k - \omega_z y_k \right), \\
    y_{k+1} &= y_k + \Delta t \left( \omega_z x_k - \frac{\Gamma}{2}y_k \right), \\
    z_{k+1} &= z_k + \Delta t \left( -\Gamma [z_k - z_{eq}(\omega_z, T)] \right),
\end{align}
where $z_{eq}(\omega_z, T) = -\tanh(\omega_z / 2T)$ defines the instantaneous thermal target.

\section{Topological Data Analysis in Phase Space}
\label{sec:tda}

The TEM workflow consists of five consecutive steps: (i) record the scalar trajectory $x(t)=\langle\sigma_x(t)\rangle$; (ii) reconstruct its phase-space geometry using time-delay embedding; (iii) compute the $H_1$ persistent homology of the resulting point cloud; (iv) represent the topological information using scalar diagram distances, persistence images, or persistence silhouettes; and (v) use these features for degradation classification and comparison with the statistical baseline. The following subsections describe each step in this order.

\subsection{Phase space reconstruction via Takens' embedding}
In experimental settings, continuously measuring the full density matrix via quantum state tomography is prohibitive. We therefore assume access to an idealized continuous record of a single local observable, $x(t)=\langle\sigma_x(t)\rangle$. According to Takens' embedding theorem \cite{takens1981}, the topology of the full multidimensional phase space can be reconstructed from a single scalar time series by constructing delay vectors
\begin{equation}
    \mathbf{v}(t) =
    \big(x(t),x(t+\tau_{\mathrm{emb}}),\dots,
    x(t+(d-1)\tau_{\mathrm{emb}})\big)^T .
    \label{eq:delay_embed}
\end{equation}
where $d$ is the embedding dimension and $\tau_{\mathrm{emb}}$ is the delay measured in sampling intervals.
We selected $\tau_{\mathrm{emb}}=10$ empirically as an intermediate delay that avoids nearly redundant coordinates at very short delays while retaining sufficient samples within the five-cycle observation window. A local sensitivity analysis using $\tau_{\mathrm{emb}}=8$, $10$, and $12$ yielded persistence-image AUC values of $0.8857$, $0.8782$, and $0.8603$, respectively, showing that moderate changes in the delay do not alter the main conclusion. The asymptotic nominal engine trajectory is a periodic limit cycle with intrinsic manifold dimension $m=1$. Takens' generic criterion $d\geq2m+1$ therefore gives the minimal embedding dimension $d=3$ \cite{takens1981}. A local sensitivity analysis using $d=3$, $4$, and $5$ yielded persistence-image AUC values of $0.8782$, $0.8355$, and $0.8540$, respectively, confirming that the main conclusion is preserved. The reconstructed dynamics therefore form the three-dimensional point cloud $\mathcal{P}=\{\mathbf{v}(t_i)\}$.

\subsection{Persistent homology and topological metrics}
To analyze the geometry of the reconstructed phase space, we compute its persistent homology. We construct a sequence of Vietoris-Rips simplicial complexes $\mathrm{VR}(\mathcal{P}, \epsilon)$ by connecting any two points in $\mathcal{P}$ that are separated by a distance less than a filtration scale $\epsilon$. As $\epsilon$ increases, topological features (such as connected components $H_0$ and loops $H_1$) appear (birth, $\epsilon_b$) and eventually fill in (death, $\epsilon_d$). 

We focus on the first homology group, $H_1$. In a perfectly periodic engine, the dynamics map to a single robust limit cycle, represented by a singular point in the persistence diagram with a large lifetime (persistence), $p_j = \epsilon_d^{(j)} - \epsilon_b^{(j)}$. Under control degradation, this primary cycle fragments, and numerous micro-loops emerge from irregular phase trajectories.

We quantify the global degradation by computing the distance between the observed persistence diagram $D$ and a nominally perfect reference diagram $D_\text{ref}$. We employ the 1-Wasserstein distance, which integrates all geometric distortions, and the Bottleneck distance, which captures the maximal single deviation~\cite{edelsbrunner2008, carlsson2009}:
\begin{align}
    W_1(D, D_\text{ref}) &= \inf_\gamma \sum_{x \in D} ||x - \gamma(x)||_1, \label{eq:wasserstein} \\
    B(D, D_\text{ref}) &= \inf_\gamma \sup_{x \in D} ||x - \gamma(x)||_\infty, \label{eq:bottleneck}
\end{align}
where $\gamma$ denotes a bijection between the diagrams. Using these complementary measures, we define a scalar control-quality index
\begin{equation}
\mathrm{QI} =
W_1(D, D_{\mathrm{ref}}) + B(D, D_{\mathrm{ref}}),
\label{eq:QI}
\end{equation}
which quantifies the overall topological deviation of an observed engine cycle from the ideal reference. We note that the definition of $\mathrm{QI}$ is not unique, as different distance metrics emphasize different types of degradation: the bottleneck distance highlights isolated, large-scale distortions, whereas the Wasserstein distance is more sensitive to subtle, distributed changes.

\subsection{High-dimensional vectorization}

To interface variable-length persistence diagrams with supervised machine learning algorithms, we map them into fixed-dimensional vector spaces. 

Persistence images $I(b,p)$ are constructed by transforming each birth–death pair $(b_i,d_i)$ into birth–persistence coordinates~\cite{adams2017}
\begin{equation}
(b_i,p_i), \qquad p_i = d_i - b_i ,
\end{equation}
and placing a persistence-weighted Gaussian kernel at each point \cite{adams2017}
\begin{equation}
\rho(b,p) = \sum_i w(p_i)
\exp\!\left(
-\frac{(b-b_i)^2 + (p-p_i)^2}{2\sigma^2}
\right).
\end{equation}
The persistence image pixel value is then obtained by integrating this surface over each grid cell
\begin{equation}
I_{m,n} = \int_{b_m}^{b_{m+1}} \int_{p_n}^{p_{n+1}}
\rho(b,p)\,db\,dp ,
\end{equation}
yielding a fixed-resolution image representation of the persistence diagram suitable for machine learning (See Appendix~\ref{app:silhouettes} for persistence silhouettes).

For each trajectory, the resulting $40\times40$ persistence image is flattened in a fixed row-wise order into the feature vector $\mathbf{f}_{\mathrm{PI}}\in\mathbb{R}^{1600}$. Likewise, the persistence silhouette evaluated on the $100$-point filtration grid forms the feature vector $\mathbf{f}_{\mathrm{sil}}\in\mathbb{R}^{100}$. Within each cross-validation split, every feature is standardized using only the mean and standard deviation of the training fold, and the same transformation is subsequently applied to the held-out fold. Consequently, no validation information enters either feature scaling or model fitting.

Persistence images were evaluated on a fixed $40\times40$ grid. Because the birth--persistence domains differ among the degradation models, the Gaussian width was selected relative to the corresponding coordinate scale: $\sigma_{\mathrm{PI}}=0.02$ was used for the timing-jitter and adiabatic ramp-distortion models, whereas $\sigma_{\mathrm{PI}}=0.005$ was used for the OU, high-frequency-ripple, and combined-noise models. In both cases, the width is approximately one grid interval, providing comparable smoothing in pixel units while preserving scale-resolved structures. Representative checks for the timing-jitter and OU models confirmed that moderate changes in $\sigma_{\mathrm{PI}}$ do not alter the principal conclusions. Grid resolutions of $30\times30$, $40\times40$, and $50\times50$ yielded persistence-image AUC values of $0.8695$, $0.8782$, and $0.8817$, respectively. We retained the $40\times40$ grid as a balance between spatial detail and feature dimensionality.
Persistence silhouettes were evaluated on a one-dimensional grid with $N_{\mathrm{grid}}=100$ and weighted using $w(p_j)=p_j^{1/2}$ (see Appendix~\ref{app:silhouettes} for more details). Values of $N_{\mathrm{grid}}=75$, $100$, and $125$ yielded AUC values of $0.7938$, $0.7974$, and $0.8029$, respectively. Weighting exponents $\beta=0$, $1/2$, and $1$ yielded AUC values of $0.7887$, $0.7974$, and $0.7995$, respectively. These small variations confirm that the conclusions are insensitive to the silhouette-grid resolution and weighting exponent. We retain $\beta=1/2$ as a balanced vectorization choice rather than selecting it solely to maximize classification performance.

\subsection{Machine learning and interpretability}
\label{sec:ml_methods}

We deploy a logistic regression classifier, trained on the vectorized topological features, to model the probability of degraded operation as
\begin{equation}
    p(y=1 \mid \mathbf{f}) = \sigma(\mathbf{w}^T \mathbf{f} + b),
\end{equation}

where $\sigma(x)=(1+e^{-x})^{-1}$ is the logistic sigmoid function, $y=1$ denotes degraded operation, $\mathbf{f}$ is a feature vector, and $\mathbf{w}$ and $b$ are the learning parameters.
Logistic regression was selected deliberately as a transparent linear classifier rather than as a performance-maximizing black-box model. Its coefficients provide a direct signed measure of how each persistence-image pixel or silhouette coordinate contributes to the degradation probability, which is consistent with the pixel-wise Pearson interpretability analysis. It also introduces substantially fewer modelling choices than kernel support-vector machines, random forests, or neural networks and is therefore appropriate for testing whether the topological representation itself contains discriminative physical information. The objective here is consequently not to benchmark classifier architectures, but to evaluate the diagnostic content of the TDA-derived features using a simple and reproducible decision model.
Classifier performance was evaluated using five-fold cross-validation. A local sensitivity analysis with $K=3$, $5$, and $10$ folds yielded persistence-image AUC values of $0.8762$, $0.8782$, and $0.8782$, respectively. The negligible variation confirms that the reported performance is insensitive to the choice of $K$. We retained $K=5$ as a standard compromise between training-set size, validation-set size, and computational cost. The logistic-regression classifier used standard $L_2$ regularization with inverse regularization strength $C=1$. A local sensitivity analysis using $C=0.1$, $1$, and $10$ yielded persistence-image AUC values of $0.8677$, $0.8782$, and $0.8932$, respectively. The topological classifier therefore remains effective over two orders of magnitude in regularization strength. We retained the standard untuned value $C=1$ to avoid post-hoc hyperparameter optimization on the reported dataset. In continuous noise models, the transition from coherent to degraded dynamics manifests as a gradual geometric deformation. To test our topological features against this physical ambiguity, we sample each noise scaling parameter uniformly across its operational range and define a strict, zero-buffer decision boundary at the midpoint (e.g., nominal for $\sigma_\tau \le 0.125$ and degraded for $\sigma_\tau > 0.125$). By requiring the classifier to analyze trajectories immediately adjacent to these thresholds, we eliminate artificial data gaps and retain the genuine geometric overlap associated with the continuous degradation crossover. To verify that the classifier learns underlying physics rather than data artifacts, we compute the Pearson correlation $r$ between the applied noise amplitude and individual pixel activations in the persistence images. This yields a 2D spatial heatmap that cleanly isolates the geometric coordinates driven by nonadiabatic fluctuations. 

To demonstrate the practical advantages of TEM, we compare its performance against a traditional SSM, described in the following section.

\subsection{Spectral-statistical monitor}
\label{sec:baseline}
To validate the effectiveness of the TEM, we benchmark its performance against the SSM. For each steady-state time series measurement $x(t)$, we extract a combined feature vector of six standard signal statistics, detailed in Appendix~\ref{app:ssm}. These features, derived from both the time and frequency domains, capture macroscopic amplitude distortions and distributional shape changes in the signal.
While simple geometric perturbations can be detected by these combined statistics, multi-scale phase desynchronization and internal micro-loop generation can substantially reduce the sensitivity of macroscopic statistical metrics. By evaluating this multi-feature SSM across progressively structured and localized perturbations, we identify the regimes in which high-dimensional topological tracking provides a marked advantage.
\begin{figure*}[t!]
    \centering
    \subfloat[]{
    \includegraphics[scale=0.396]{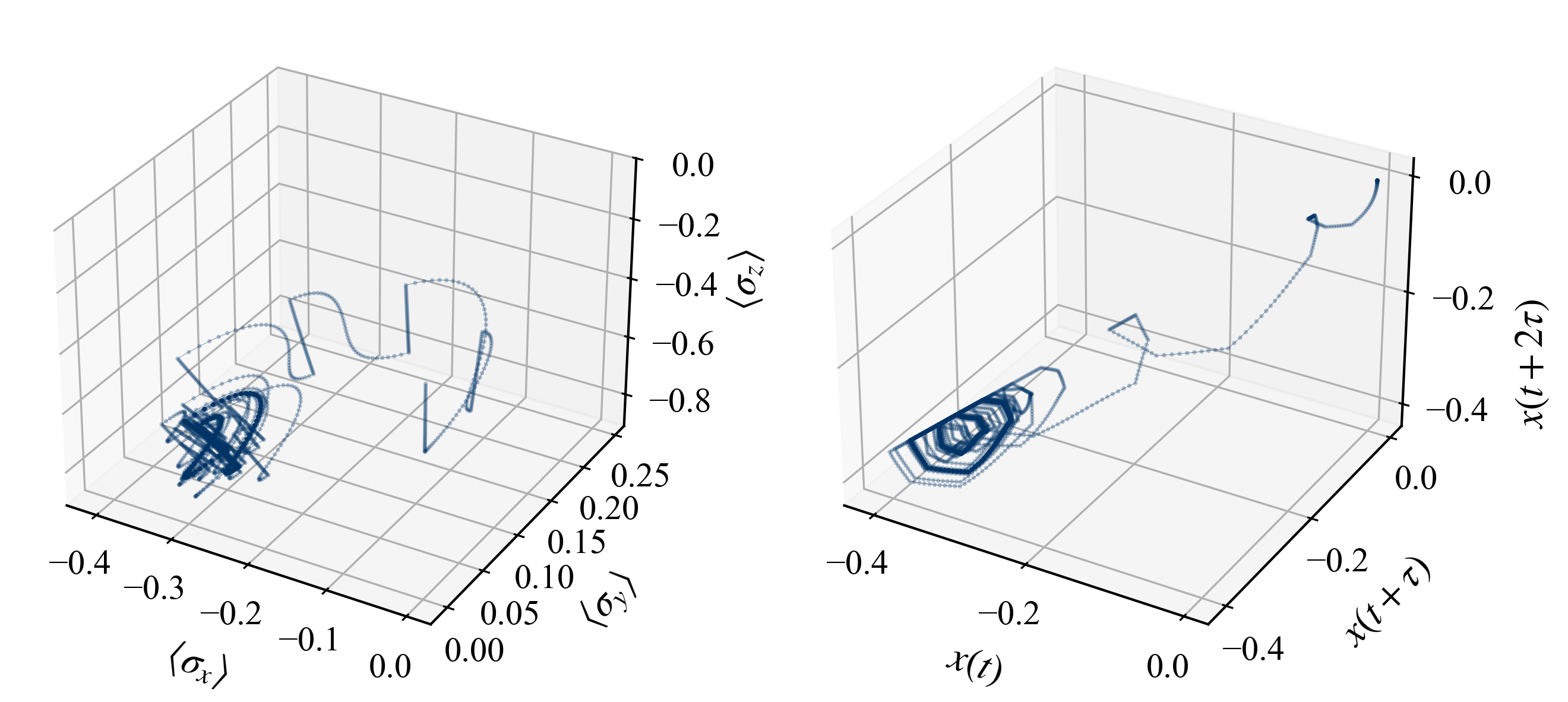}}
    \subfloat[]{
    \includegraphics[scale=0.396]{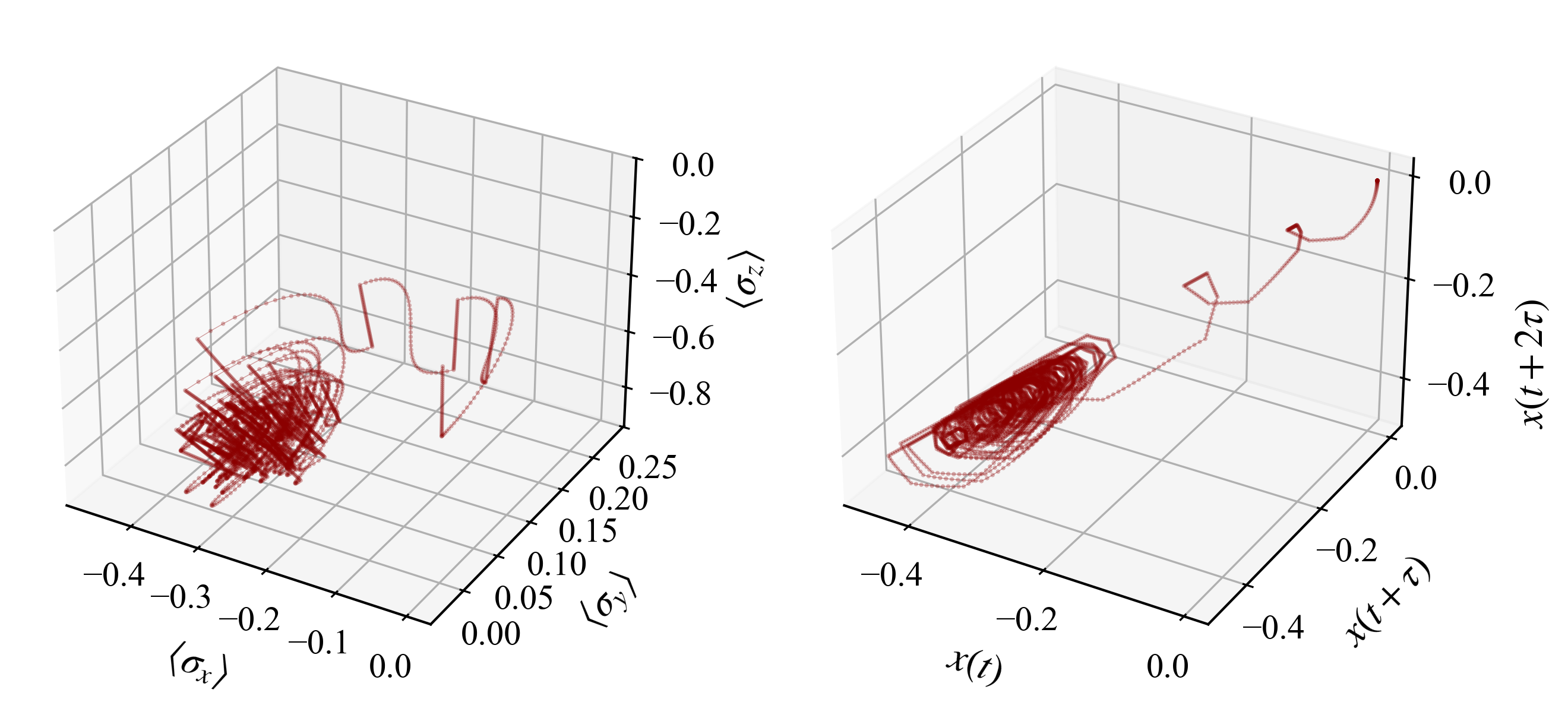}}\\
    \subfloat[]{
    \includegraphics[scale=0.7]{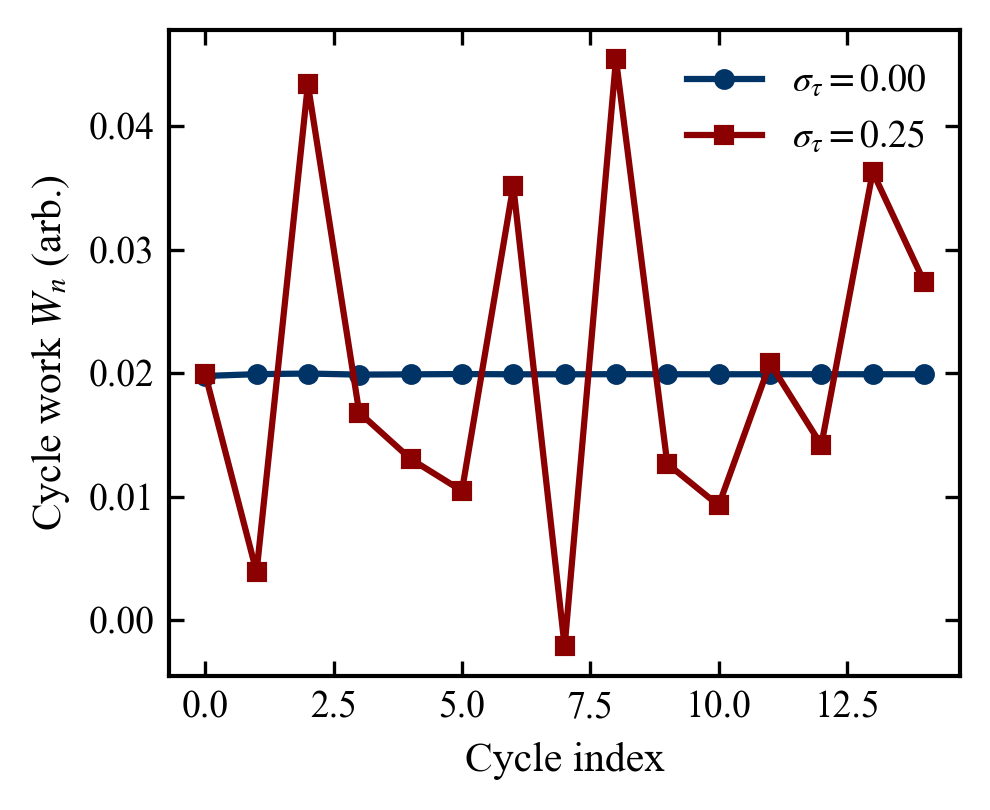}\label{fig:cyclework_a}}
    \subfloat[]{
    \includegraphics[scale=0.7]{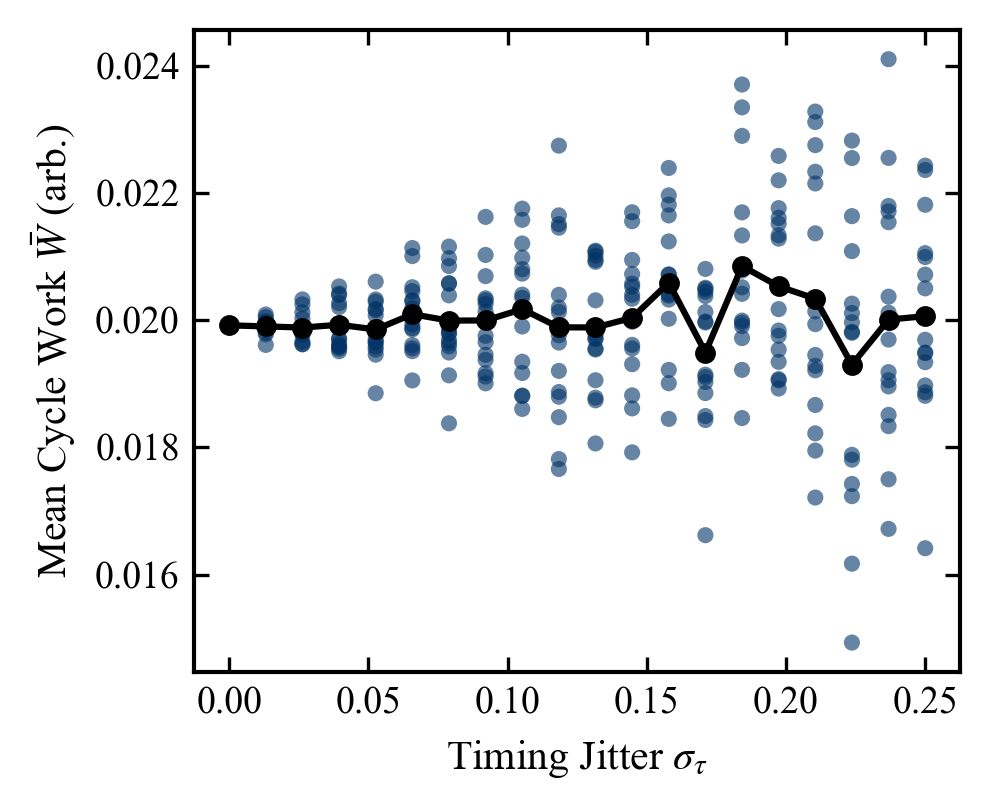}\label{fig:cyclework_b}}
    \subfloat[]{
    \includegraphics[scale=0.7]{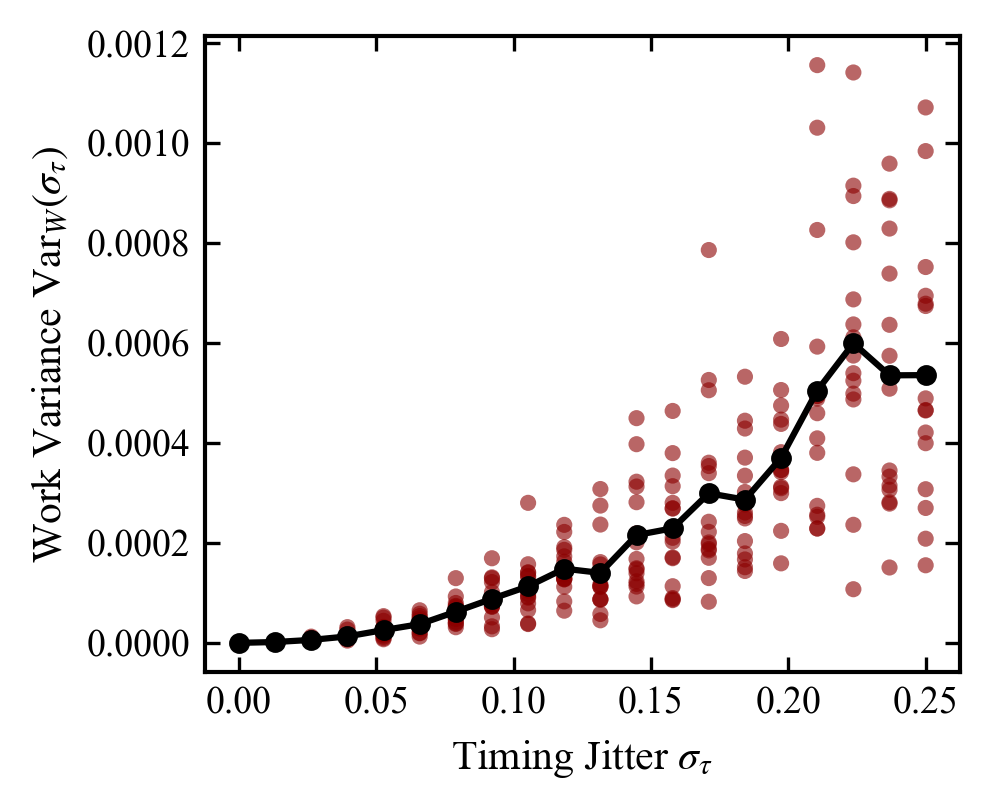}\label{fig:cyclework_c}}
    \caption{Phase space trajectories of the quantum Otto cycle. (a) In the nominal regime ($\sigma_\tau = 0$), the engine settles into a stable, well-defined limit cycle in both the physical 3D Bloch sphere (left) and the corresponding time-delay embedding (right). (b) Under extreme timing jitter ($\sigma_\tau = 0.25$), nonadiabatic friction destroys the strict periodicity of the limit cycle, producing a stochastically broadened geometric manifold in both representations. (c) Instantaneous cycle work $W_n$ as a function of cycle index $M$. The blue circles represent stable nominal operation ($\sigma_\tau = 0.00$), while the red squares show irregular amplitude fluctuations under extreme timing jitter ($\sigma_\tau = 0.25$). (d) Mean cycle work $\bar{W}$ and (e) work variance $\mathrm{Var}_W$ as a function of timing jitter $\sigma_\tau$. In both (d) and (e), individual operational runs are plotted as transparent scatter points (blue for $\bar{W}$, red for $\mathrm{Var}_W$), while the solid black line with circular markers indicates the ensemble average at each discrete noise level.}
    \label{fig:bloch_delay}
\end{figure*}
\section{Results and Discussion}
\label{sec:results}
Here, we analyze the engine's response to control degradation using both traditional thermodynamic observables and topological diagnostics. We first examine the behavior of energetic quantities such as instantaneous and cycle-averaged work, and then introduce the TDA methods to reveal the structural breakdown of the engine dynamics. We structure
this analysis across five progressively structured and localized
control-perturbation settings to compare the diagnostic performance of TEM and SSM.

To obtain the results presented in this section, the Bloch equations were numerically integrated using a discretized Euler scheme over successive engine cycles (see Sec.~\ref{sec:numerical_integration}).
The engine parameters were fixed at $\omega_h=2.0$, $\omega_c=1.0$, $T_h=0.8$, $T_c=0.25$, $\omega_x^{\max}=1.0$, and $\Gamma=0.6$, with nominal stroke durations $\tau_h=\tau_c=0.7$ and $\tau_1=\tau_3=0.6$. The corresponding ideal Otto cycle operates in the heat-engine regime, which requires $\omega_h/T_h<\omega_c/T_c$ \cite{kosloff2013,Rezek_2006}; specifically, $\omega_h/T_h=2.5<4.0=\omega_c/T_c$.

The transverse-drive amplitude $\omega_x^{\max}=1.0$ places the transverse and longitudinal energy scales at comparable magnitudes. Values of $\omega_x^{\max}=0.8$, $1.0$, and $1.2$ yielded persistence-image AUC values of $0.8838$, $0.8782$, and $0.8786$, respectively, confirming insensitivity to moderate changes in the drive amplitude.

The relaxation rate $\Gamma=0.6$ corresponds to the dimensionless isochore relaxation strength $\Gamma\tau_h=0.42$. Values of $\Gamma=0.4$, $0.6$, and $0.8$ yielded persistence-image AUC values of $0.8866$, $0.8782$, and $0.8981$, respectively, confirming stability across this relaxation range.

The isochoric-stroke durations were fixed at $\tau_h=\tau_c=0.7$. Values of $\tau_h=\tau_c=0.6$, $0.7$, and $0.8$ yielded persistence-image AUC values of $0.8921$, $0.8782$, and $0.8862$, respectively, indicating insensitivity to moderate changes in the isochore duration.
\begin{figure*}[t]
    \centering
    \includegraphics[scale=0.94]{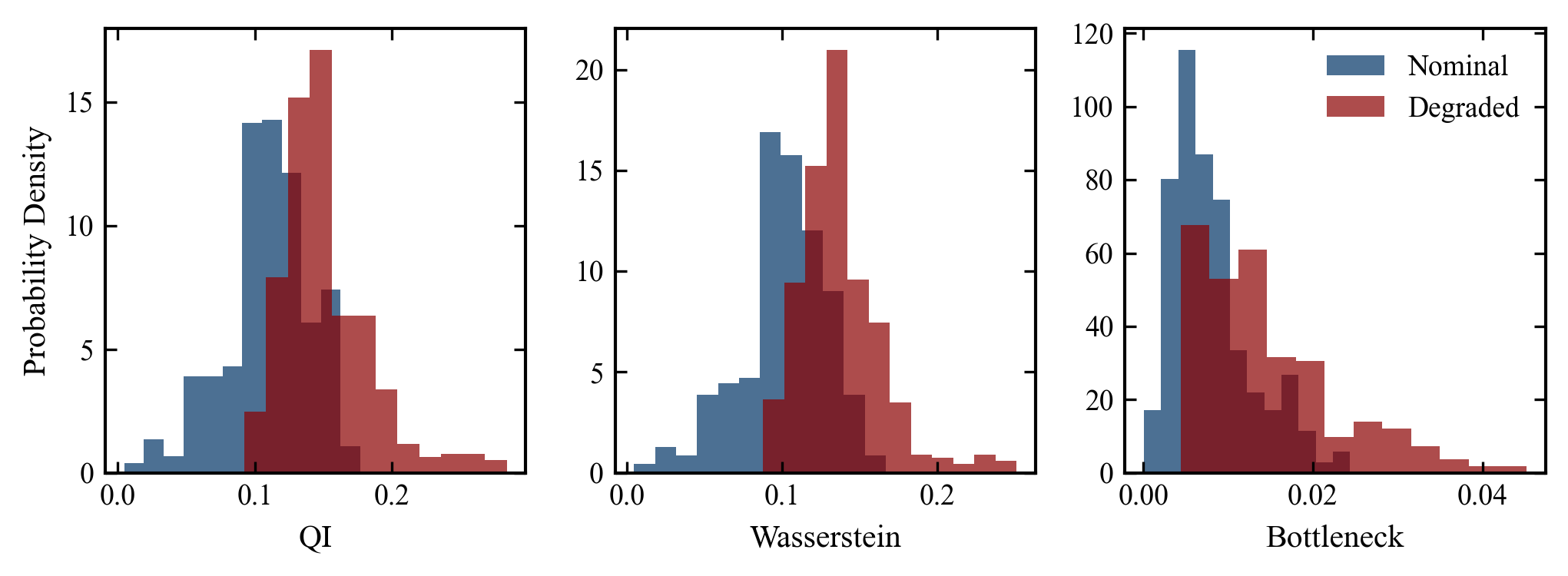}
    \caption{Probability density histograms of three TDA-based metrics—QI, Wasserstein distance, and Bottleneck distance—for nominal (blue) and degraded (orange) engine conditions, showing systematic distributional shifts with partial overlap between the two regimes.}
    \label{fig:histograms}
\end{figure*}
The unitary-stroke durations were fixed at $\tau_1=\tau_3=0.6$. Values of $\tau_1=\tau_3=0.5$, $0.6$, and $0.7$ yielded persistence-image AUC values of $0.8646$, $0.8782$, and $0.9123$, respectively. Thus, the absolute performance varies with the available coherent-evolution time, but the topological monitor remains effective throughout the tested range. All central physical parameters were retained without post-hoc performance optimization. Crucially, we employ distinct sampling windows to separate physical visualization from diagnostic classification. To ensure the visualizations accurately reflect the steady-state thermodynamics of the limit cycle, an extended transient \textit{burn-in} period of $15$ cycles was discarded---allowing the system to fully relax and lose memory of its arbitrary initial state---with data collected over the subsequent $M=15$ cycles. For the visualization of thermodynamic observables under timing jitter, we generated 15 independent trajectories across a discrete 20-point grid up to a maximum jitter amplitude of $\sigma_\tau = 0.25$. However, to evaluate the algorithms in rapid-response, early-degradation detection scenarios, the machine-learning datasets were constructed using a burn-in of two cycles followed by an evaluation window of $M=5$ cycles, corresponding to $800$ sampled trajectory points. Burn-in durations of $1$, $2$, and $3$ cycles resulted in persistence-image AUC values of $0.8969$, $0.8782$, and $0.8668$, respectively, indicating that the main conclusions are insensitive to the precise transient-removal interval. We retained two burn-in cycles as a compromise between reducing dependence on the arbitrary initial state and enabling rapid detection. A separate observation-window sensitivity analysis with $M=3$, $4$, and $5$ cycles produced AUC values of $0.8267$, $0.8440$, and $0.8782$, respectively. We therefore retained $M=5$, which provides the highest and most stable performance while requiring only five post-burn-in engine cycles.  This intentional inclusion of transient dynamical drift requires the classifiers to contend with transient, non-asymptotic engine operation. To systematically probe the continuous degradation crossover across all five perturbation settings, we generated $1000$ independent trajectories per degradation model. The corresponding noise-intensity parameter was sampled uniformly across its operational range, producing approximately balanced nominal and degraded classes. A stratified dataset-size sensitivity analysis using $500$, $750$, and $1000$ trajectories gave persistence-image AUC values of $0.8635$, $0.8727$, and $0.8775$, respectively. The gradual saturation of the AUC indicates convergence with increasing sample size; therefore, $1000$ trajectories were retained for the reported analysis. For all dynamical analyses, delay embeddings were constructed using an embedding dimension of $d=3$ and a delay of $\tau_{\mathrm{emb}}=10$ sampling intervals.

\subsection{Thermodynamic observables of the engine dynamics}
\label{sec:energetic_diagnostics}

The physical impact of finite-time driving and timing jitter is first visualized in the operational phase space, as presented in Fig.~\ref{fig:bloch_delay}. To illustrate the topological transformation of the engine, Figs.~\ref{fig:bloch_delay}(a) and (b) contrast the physical 3D Bloch vector trajectory (left panels) with its corresponding mathematical time-delay embedding (right panels).
Under ideal nominal control ($\sigma_\tau = 0$, Fig.~\ref{fig:bloch_delay}(a)), the engine synchronizes with the periodic drive. The Bloch vector $\mathbf{r}(t)$ establishes a stable, well-defined limit cycle, reflecting the perfect balance between the coherent unitary rotations of the work strokes and the Markovian dissipation of the thermal isochores. When this scalar observable $x(t) = \langle\sigma_x(t)\rangle$ is mapped into the reconstructed phase space using Takens' embedding (Eq.~\eqref{eq:delay_embed}), the limit cycle is perfectly preserved as a sharp, continuous 1D geometric loop.

However, when strong control degradation is introduced ($\sigma_\tau=0.25$, Fig.~\ref{fig:bloch_delay}(b)), the periodicity of the drive is lost. Nonadiabatic quantum friction and randomized phase accumulation prevent the qubit from returning to the same state at the end of each cycle. In the physical Bloch sphere, the sharp limit cycle is replaced by a stochastically broadened set of trajectories. The time-delay embedding inherits this loss of periodicity: the previously sharp topological loop becomes a highly self-intersecting, diffuse manifold. TEM is designed to quantify this smooth geometric deformation from nominal cyclic operation toward increasingly degraded dynamics.

Figures~\ref{fig:bloch_delay}(c)--(e) demonstrate the limitations of thermodynamic tracking. As shown in Fig.~\ref{fig:bloch_delay}(c), while the work output $W_n$ is relatively stable under nominal conditions, strong timing jitter causes large, irregular fluctuations and can produce cycles with net-negative work. Furthermore, Fig.~\ref{fig:bloch_delay}(d) shows that the mean cycle work $\bar{W}$ is non-monotonic and highly variable across the degradation range, providing no clear boundary for degradation detection. Although the work variance in Fig.~\ref{fig:bloch_delay}(e) increases with degradation, its calculation requires extensive temporal averaging, preventing short-window, trajectory-level diagnosis.

Traditional diagnostics of engine performance rely on energetic observables such as work, heat exchange, and efficiency. Although these quantities are physically meaningful, we observe that they exhibit strong fluctuations in finite-time engines, even under nominal operating conditions. As a result, they are not always well-suited for automated monitoring, identification of dynamical irregularities, or model-independent assessment of control quality. This motivates the search for alternative diagnostics capable of capturing the underlying dynamical structure of the engine cycle in a more robust manner. Therefore, in the next section, we discuss the performance of the TEM framework and compare it with the SSM baseline.

\subsection{Global cycle corruption}
\label{sec:global_corruption}
We first investigate the capacity of scalar topological metrics to distinguish nominal from degraded engine operation under global cycle corruption. Under this macroscopic degradation model, the duration $\tau_j$ of each operational stroke is independently perturbed by uncorrelated Gaussian timing jitter (Eq.~\eqref{eq:timing_jitter}), representing a severe global clock desynchronization. 

Figure~\ref{fig:histograms} presents the probability distributions of the scalar topological distances, partitioned at the classification threshold $\sigma_\tau=0.125$. Unlike the strongly fluctuating thermodynamic variables discussed previously, the topological metrics exhibit structured distributions, with a narrow overlap near the classification boundary.
\begin{figure}[t!]
    \centering
    \includegraphics[scale=0.94]{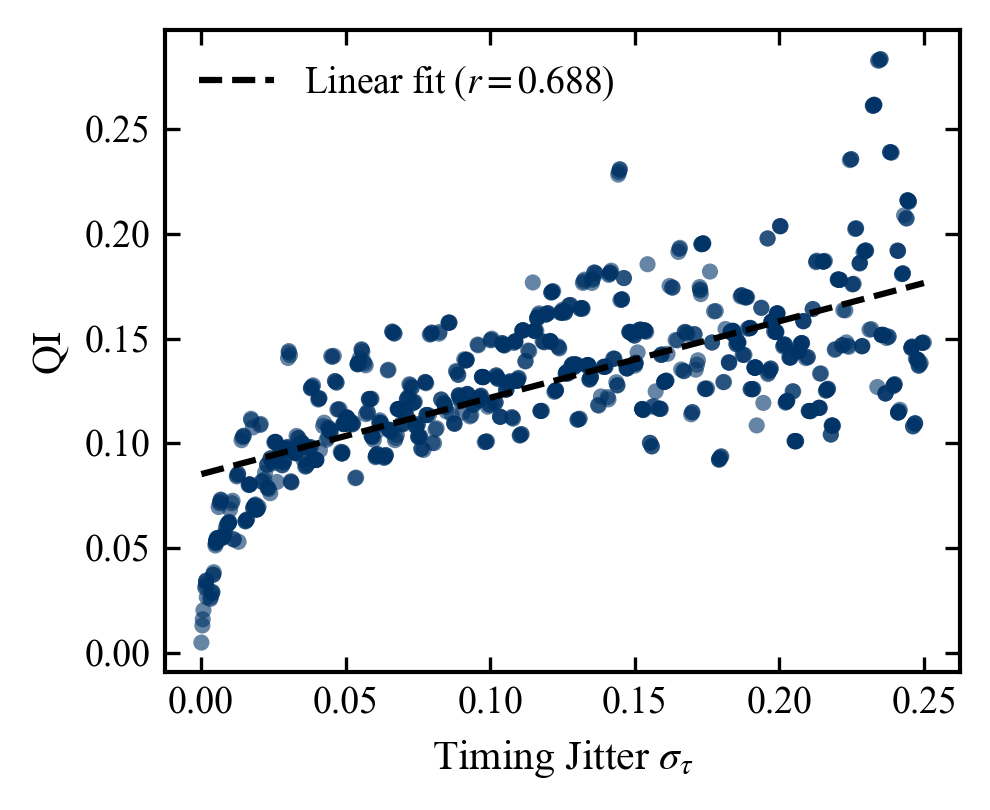}
    \caption{QI as a function of timing jitter $\sigma_\tau$. The linear correlation ($r=0.688$) tracks cumulative geometric degradation across the continuous noise spectrum, confirming a smooth structural breakdown of the engine limit cycle.}
    \label{fig:qi_scatter}
\end{figure}
Building on this, Fig.~\ref{fig:qi_scatter} illustrates the QI scaling linearly with timing jitter ($r=0.688$). The continuous sampling of $\sigma_\tau$ reveals a smooth geometric degradation crossover between the nominal and degraded regimes rather than a discrete transition.

To capture the full geometric structure beyond scalar metrics, we map the persistence diagrams into high-dimensional feature spaces.
\begin{figure}[b]
    \centering
    \includegraphics[scale=0.5]{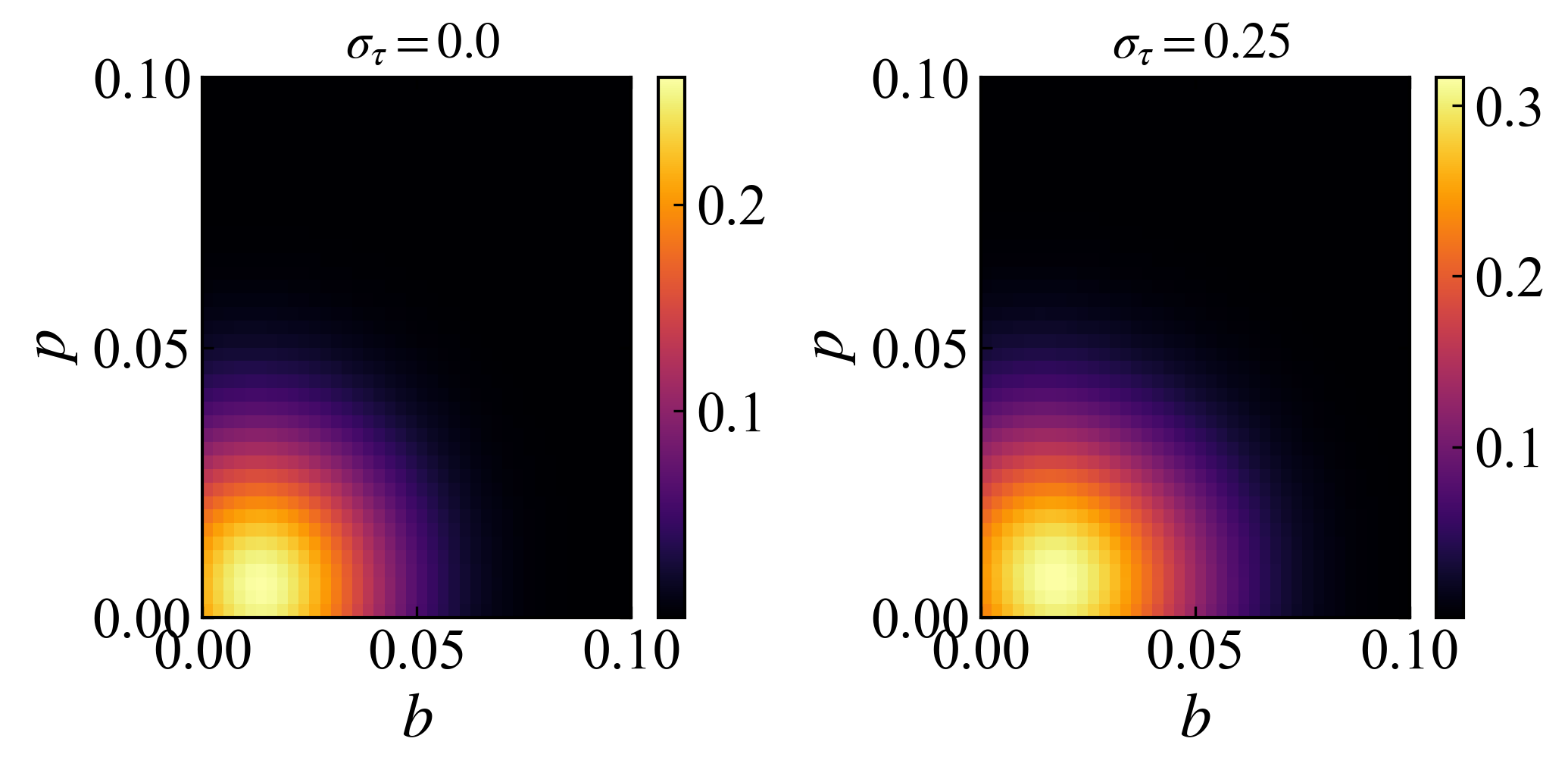}
    \caption{Persistence images comparing cases with zero timing jitter (left) and extreme timing jitter (right). The timing noise smears the primary geometric invariants into diffuse topological noise.}
    \label{fig:vectorizations}
\end{figure}
As shown in Fig.~\ref{fig:vectorizations}, extreme jitter forces the collapse of localized geometric hotspots in the persistence images and gives rise to diffuse, low-persistence noise artifacts. The logistic-regression models trained on these topological maps use these features for automated degradation detection.

\begin{figure}[t]
    \centering
    \includegraphics[scale=0.94]{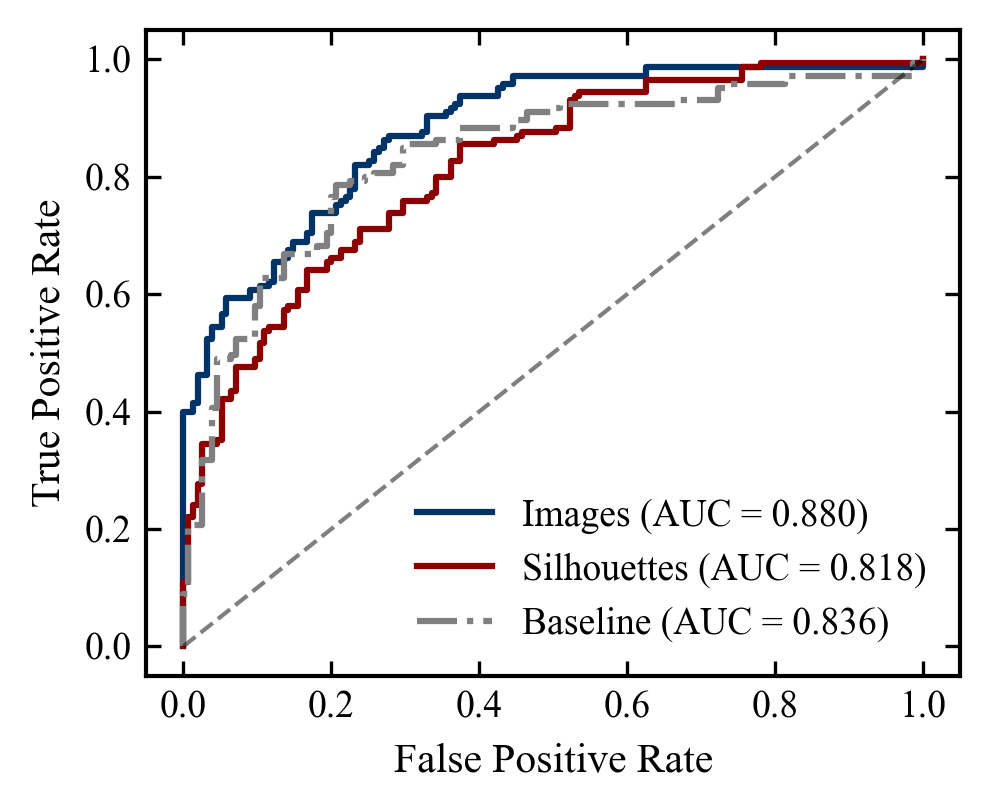}
    \caption{ROC curves for binary classification of nominal ($\sigma_\tau \le 0.125$) versus degraded ($\sigma_\tau > 0.125$) engine operation across the continuous noise spectrum, using persistence images (solid blue curve) and persistence silhouettes (solid red curve) as topological feature vectors. Curves are shown for a single representative train-test split, whereas the summarized performance metrics in Table I correspond to the 5-fold cross-validated mean.}
    \label{fig:roc}
\end{figure}

As shown in Fig.~\ref{fig:roc}, the receiver operating characteristic (ROC) curve characterizes the binary classification performance across the continuous noise spectrum, distinguishing nominal ($\sigma_\tau \le 0.125$) from degraded ($\sigma_\tau > 0.125$) engine operation. We compare two topological feature representations—persistence images (solid blue curve) and persistence silhouettes (solid red curve)—against the SSM (dashed gray curve). By eliminating the training buffer zone, the classifier is required to discriminate directly across the classification boundary, yielding non-perfect performance that reflects the smooth overlap between nominal and degraded trajectories. Persistence images achieve the highest area under the curve ($\mathrm{AUC}=0.8782$), successfully outperforming both the strong SSM ($\mathrm{AUC}=0.8360$) and the persistence silhouettes ($\mathrm{AUC}=0.7974$). This performance hierarchy physically aligns with the macroscopic nature of global timing jitter: because this specific noise severely alters the gross phase-space volume and cycle duration, the combined SSM metrics compete highly effectively. While the 1D skyline projection of the silhouettes loses some of this volumetric data, the full 2D spatial representation of persistence images successfully resolves both the macroscopic expansion and the subtle, overlapping topological deformations near the classification boundary. The dashed diagonal line denotes random chance ($\mathrm{AUC}=0.5$).

\begin{figure}[t]
    \centering
    \includegraphics[scale=1.02]{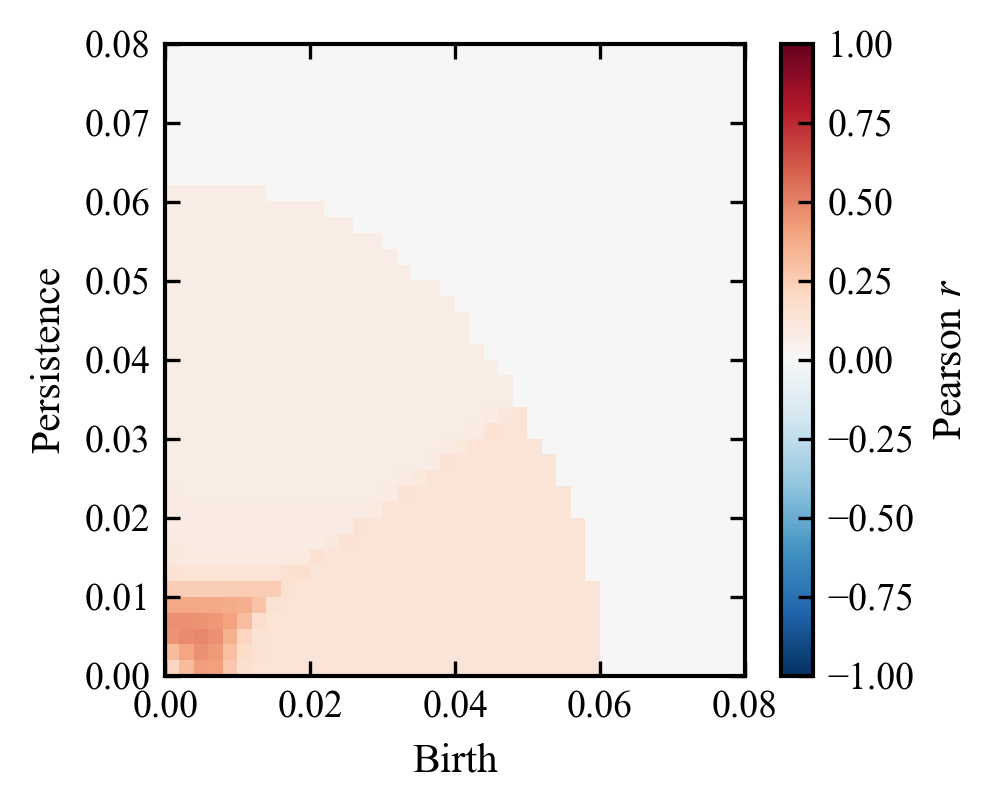}
    \caption{Pixel-wise Pearson correlation heatmap. The positive correlations (dark red) localized near the origin confirm that quantum friction manifests geometrically as an accumulation of high-frequency, low-persistence microscopic loops.}
    \label{fig:heatmap}
\end{figure}

Finally, Fig.~\ref{fig:heatmap} provides a physical interpretation of the machine learning model. The spatial Pearson correlation heatmap reveals that the deep positive correlations (dark red) strictly localize near the origin (low Birth, very low Persistence). This confirms that timing jitter does not uniformly expand the phase space; rather, quantum friction generates highly localized, microscopic self-intersecting loops. The model isolates these nonadiabatic-friction signatures to identify degraded operation. However, because this global duration jitter severely alters the macroscopic length and overall volume of the trajectory loop, the non-topological SSM retains competitive classification performance, capturing these macroscopic distortions (Table~\ref{tab:auc_scores}). In this regime, global jitter acts as a coarse perturbation that is detectable by both geometric and scalar statistical measures. Therefore, we turn to more structured and localized perturbation models in the following sections.

\subsection{Beyond Gaussian timing jitter}
\label{sec:beyond_jitter}
To systematically isolate the discriminative capabilities of TEM, we progressively introduce the localized, physically motivated benchmark perturbations defined in Sec.~\ref{sec:degradation_models}. While global timing jitter acts macroscopically on the overall volume of the phase space, the following sub-sections examine the robustness of both our topological framework and the SSM against specific geometric sweep deformations, finite-bandwidth colored noise, phase-coherent perturbations, and combined multi-channel degradation.
\begin{table*}[t]
\centering
\setlength{\tabcolsep}{30pt} 
\caption{Classification performance (AUC) of the spectral-statistical monitor (SSM) versus the topological engine monitor (TEM) using persistence images and silhouettes across progressively structured control-degradation models.}
\label{tab:auc_scores}
\begin{tabular}{l c c c}
\hline\hline
\textbf{Degradation Model} & \textbf{SSM} & \textbf{TEM-Images} & \textbf{TEM-Silhouettes} \\
\hline
Global cycle corruption & 0.8360 & 0.8782 & 0.7974 \\
Adiabatic ramp distortion & 0.7660 & 0.8537 & 0.8135 \\
Correlated sweep noise (OU) & 0.6932 & 0.9703 & 0.9490 \\
Longitudinal high-frequency ripple & 0.6759 & 0.9474 & 0.9071 \\
Combined multi-channel control degradation & 0.6817 & 0.9278 & 0.8463 \\
\hline\hline
\end{tabular}
\end{table*}

\subsubsection{Adiabatic ramp distortion}
\label{sec:ramp_distortion_results}
Under the adiabatic ramp distortion model (Eqs.~\eqref{eq:ramp_omega}--\eqref{eq:ramp_deriv}), the trajectory ramp bows inward or outward, distorting the macroscopic sweep velocity without generating internal high-frequency friction loops. Because this structural defect acts primarily on the global amplitude and skew of the dynamical trajectory, the SSM maintains a moderate detection capability (see Table~\ref{tab:auc_scores}). However, this macroscopic distortion highlights an intermediate case where the topological vectorizations retain a clear predictive advantage: the 2D spatial resolution of the TEM successfully captures the geometric bowing of the trajectory, significantly outperforming the standard time-domain statistical analysis.

\subsubsection{Correlated adiabatic sweep noise}
\label{sec:ou_noise_results}
Under the correlated adiabatic sweep noise governed by Eq.~\eqref{eq:ou_sde}, the OU process continuously mean-reverts toward the nominal value. Consequently, the perturbation largely preserves the global shape and duration of the thermodynamic cycle. The SSM therefore loses much of its discriminative sensitivity and approaches weak classification performance.

The finite bandwidth of the colored noise nevertheless generates localized, high-frequency micro-loops along the trajectory. Persistent homology captures these sub-scale geometric structures, enabling TEM to retain strong discrimination where the standard statistical moments provide only limited sensitivity (Table~\ref{tab:auc_scores}).

\subsubsection{Longitudinal high-frequency ripple}
\label{sec:coherence_ripple_results}
Under the longitudinal high-frequency ripple in Eq.~\eqref{eq:ripple_omega}, the phase-space geometry encodes the control degradation: the sinusoidal perturbation folds the limit cycle into structured topological sub-cycles. Because this even-$k$ ripple preserves the boundary conditions and general energetic envelope of the Otto cycle, the SSM shows limited discrimination between nominal and degraded cycles. In contrast, the TEM captures this emergent resonant geometry effectively, maintaining excellent classification performance (Table~\ref{tab:auc_scores}), further establishing topological indexing as the robust indicator for detecting edge-case phase desynchronizations.

\subsubsection{Combined multi-channel control degradation}
\label{sec:combined_degradation_results}
The combined multi-channel degradation model provides the most stringent benchmark considered here. The residual macroscopic trajectory skewing remains partly visible to the SSM, but its discriminative performance is reduced by the simultaneous presence of high-frequency internal structure. In contrast, the persistent-homology framework captures both the macroscopic coordinate deformation and the density of internal micro-loops. TEM, particularly through the two-dimensional persistence-image representation, therefore retains a marked advantage in this combined benchmark model (Table~\ref{tab:auc_scores}).

\section{Conclusion and Outlook}
\label{sec:conclusion}
In this work, we applied a single-observable, topology-driven framework for condition monitoring and control-degradation detection in finite-time quantum thermodynamic cycles. We demonstrated that, in the nonadiabatic driving regime, traditional energetic diagnostics such as instantaneous and mean cycle work are strongly affected by inherent quantum friction and irregular cycle-to-cycle fluctuations. Although the variance of the work output correlates with control degradation, its calculation requires extensive statistical averaging, limiting its suitability for rapid diagnostics.

To overcome these limitations, we reconstructed the phase-space dynamics of the quantum Otto engine using Takens time-delay embeddings of the observable $x(t)=\langle\sigma_x(t)\rangle$. By computing the $H_1$ persistent homology of these reconstructed manifolds, we mapped the deformation of the engine limit cycle into a quantifiable geometric representation. The scalar QI, derived from the Wasserstein and Bottleneck distances, exhibited an approximately linear dependence ($r=0.688$) on the timing-jitter magnitude. It therefore provides a systematic geometric signature of increasing control degradation before a pronounced loss of stable thermodynamic operation occurs.

We further established an automated classification pipeline by vectorizing the topological features into persistence images and silhouettes. By continuously sampling the timing jitter and evaluating trajectories directly across the classification boundary, the logistic-regression classifiers achieved robust discrimination under physically motivated benchmark perturbations. The comparison with the SSM showed that, while conventional statistical features remain effective for macroscopic trajectory deformations, their performance degrades substantially under localized, finite-bandwidth perturbations that preserve the global phase-space boundaries. In these regimes, TEM provides a marked diagnostic advantage by detecting the internal micro-loop structures associated with high-frequency quantum-friction signatures. Finally, the pixel-wise Pearson correlation analysis identified the specific low-persistence regions associated with cyclic degradation, confirming that the classifier learns physically interpretable geometric features rather than arbitrary data artifacts.

The topological framework presented here opens several promising avenues for quantum control and thermodynamics. First, while this study focused on a single-qubit working medium, TDA is natively high-dimensional. Extending this methodology to multi-qubit engines could allow for the topological monitoring of entanglement generation and many-body decoherence during finite-time strokes. Second, because persistent homology is robust against coordinate deformations, this geometric diagnostic channel could be integrated into real-time, closed-loop feedback protocols for autonomous quantum engines, allowing them to dynamically correct timing jitter on the fly. Finally, the reliance on an idealized continuous record of a single observable makes this pipeline highly amenable to near-term experimental implementations in superconducting circuit QED \cite{blais2021} or trapped-ion platforms \cite{bruzewicz2019}, where full quantum state tomography remains a prohibitive bottleneck for continuous operation.

\section{ACKNOWLEDGMENTS}
We thank Şeyda Leyla Bozan of Koç University for fruitful discussions. This work was partially supported by National Science Foundation under grants DMS-2220613, and DMS-2229417.

\section{DATA AVAILABILITY}

The code and synthetic-data generation workflows supporting the findings of this study are publicly available at~\cite{Maden2026TEMSupplement}. All benchmark trajectories are generated by the notebooks in the repository; no external dataset was used.

\appendix

\section{Baseline feature expressions}\label{app:ssm}
To provide a rigorous non-topological baseline for evaluating the discriminative power of TDA, we extract a combined feature vector of six standard signal statistics from each steady-state time-series measurement $x(t)$ of discrete length $N$. These features are designed to capture macroscopic amplitude deformations and changes in the distribution shape:

\textit{1. Standard Deviation ($\sigma$):} Quantifies the overall spread of the signal fluctuations around the mean $\mu = \frac{1}{N} \sum_{i=1}^{N} x_i$,
\begin{equation}
    \sigma = \sqrt{\frac{1}{N} \sum_{i=1}^{N} (x_i - \mu)^2}.
\end{equation}

\textit{2. Skewness ($\gamma_1$):} Measures the asymmetry of the signal's amplitude distribution,
\begin{equation}
    \gamma_1 = \frac{1}{N} \sum_{i=1}^{N} \left(\frac{x_i - \mu}{\sigma}\right)^3.
\end{equation}

\textit{3. Kurtosis ($\beta_2$):} Measures the "tailedness" of the distribution, which is highly sensitive to extreme trajectory outliers and irregular loops,
\begin{equation}
    \beta_2 = \frac{1}{N} \sum_{i=1}^{N} \left(\frac{x_i - \mu}{\sigma}\right)^4.
\end{equation}

\textit{4. Peak-to-Peak Amplitude ($x_{p-p}$):} Captures the absolute macroscopic bounds of the dynamical limit cycle,
\begin{equation}
    x_{p-p} = \max_{i}(x_i) - \min_{i}(x_i).
\end{equation}

\textit{5. Root-Mean-Square (RMS):} Represents the effective power or gross amplitude of the observable,
\begin{equation}
    x_{\mathrm{rms}} = \sqrt{\frac{1}{N} \sum_{i=1}^{N} x_i^2}.
\end{equation}

\textit{6. Spectral Centroid ($f_c$):} A frequency-domain feature that identifies the "center of mass" of the signal's spectrum. If $X(f_k)$ represents the discrete Fourier transform of $x(t)$ at frequency bin $f_k$, then:
\begin{equation}
    f_c = \frac{\sum_{k} f_k |X(f_k)|}{\sum_{k} |X(f_k)|}.
\end{equation}

\section{TDA hyperparameters and vectorization}\label{app:tda_params}
The persistent homology calculations target the $H_1$ homology group (1-dimensional cycles/loops), ignoring $H_0$ (connected components) as the phase space trajectory is continuous. 


For the Persistence Silhouettes, the features were mapped to a 1D grid with a resolution of $N_\text{grid} = 100$. The weighting function for the triangular height was explicitly chosen as $w(p_j) = \sqrt{p_j}$. While standard TDA literature often uses linear weighting ($w(p) = p$) \cite{adams2017} or constant weighting ($w(p) = 1$) \cite{chazal2014}, the square-root function was strategically selected to compress the dominant macroscopic limit cycle while simultaneously boosting the visibility of the low-persistence micro-structures generated by nonadiabatic quantum friction.

\begin{figure}[b]
    \centering
    \includegraphics[scale=0.35]{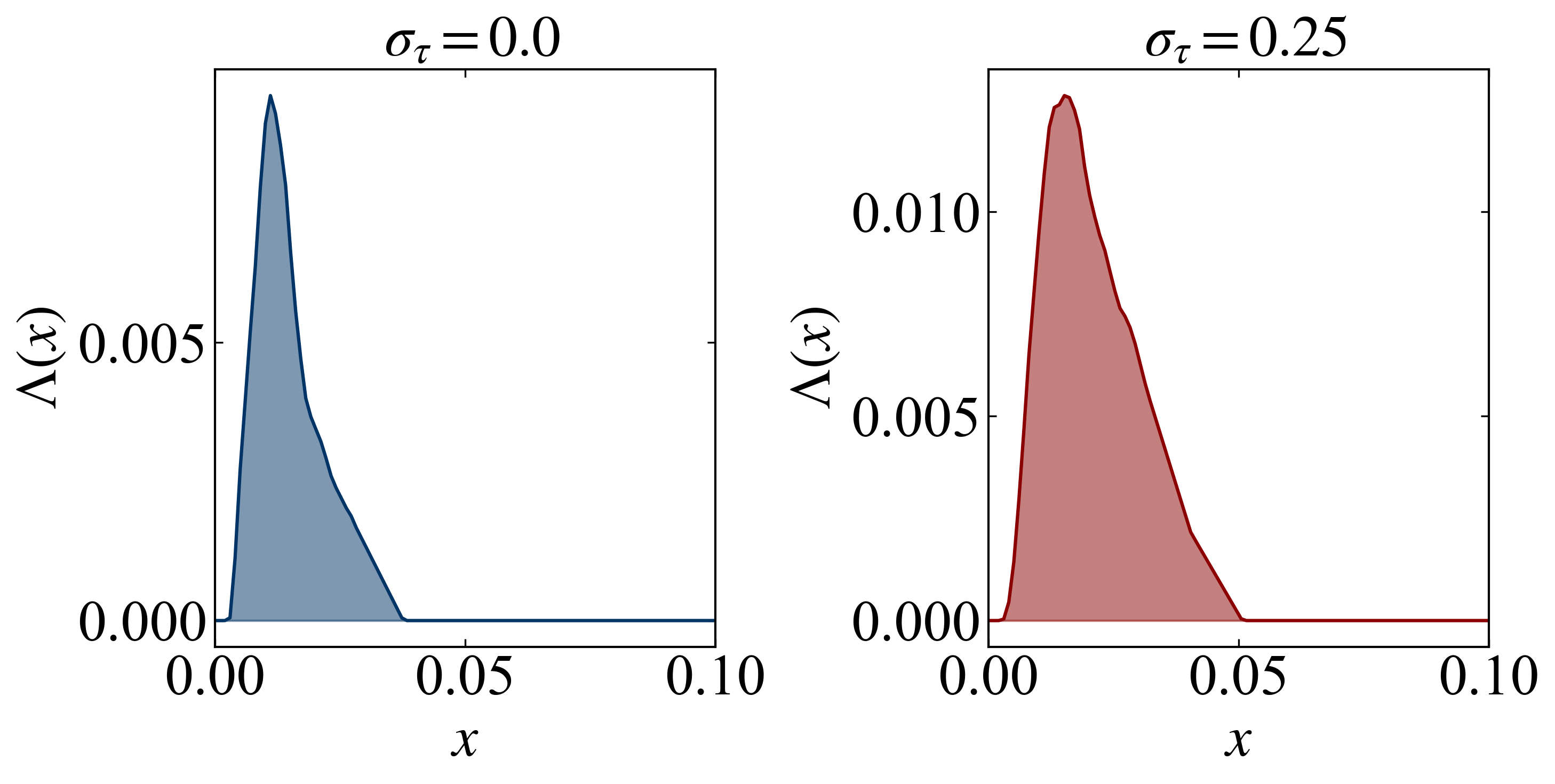}
    \caption{Silhouette height $\Lambda(x)$ as a function of filtration variable $x$ for nominal $(\sigma_{\tau}=0)$ and degraded $(\sigma_{\tau}=0.25)$ regimes.}
    \label{fig:PS}
\end{figure}

\section{Persistence silhouettes}\label{app:silhouettes}
Persistence silhouettes $\Lambda(x)$ map the features to a 1D skyline array across the filtration scale \cite{chazal2014}, which is defined as 
\begin{equation}
    \Lambda(x) = \sum_{j \in D} \sqrt{p_j} \max\left(0, \frac{p_j}{2} - \left|x - \frac{b_j+d_j}{2}\right|\right).
\end{equation}
where $\mathcal{D}$ is the persistence diagram, $p_j = d_j - b_j$ is the persistence (lifetime) of the $j$-th topological feature, and the tent function is centered at the midpoint $(b_j+d_j)/2$ of each birth–death pair. The $\sqrt{p_j}$ weighting interpolates between uniform treatment of all features ($p_j^0$) and dominance by the most persistent ones 
($p_j^1$). The $\sqrt{p_j}$ scaling enhances the relative visibility of noise-induced micro-structures that act as early indicators of cyclic degradation.
Figure~\ref{fig:PS} represents the persistence silhouette $\Lambda(x)$, as a function of filtration value $x$. In the nominal regime ($\sigma_\tau = 0$), $\Lambda(x)$ exhibits a sharp, narrow peak, 
reflecting a nearly periodic qubit trajectory whose topological features are concentrated at a single characteristic scale. Under timing noise 
($\sigma_\tau = 0.25$), the silhouette broadens and its amplitude increases by nearly an order of magnitude, signaling that the trajectory acquires topological structure across multiple scales — a direct signature of cycle deformation induced by quantum friction and stochastic driving imperfections. Since this diagnostic operates entirely at the level of the persistent homology of the reconstructed trajectory, it requires no direct access to individual Bloch vector components or operator expectation values, providing a robust, coordinate-free criterion for distinguishing nominal from degraded engine operation.

\bibliography{references}

@Article{kosloff2013,
AUTHOR = {Kosloff, Ronnie},
TITLE = {Quantum Thermodynamics: A Dynamical Viewpoint},
JOURNAL = {Entropy},
VOLUME = {15},
YEAR = {2013},
NUMBER = {6},
PAGES = {2100--2128},
URL = {https://www.mdpi.com/1099-4300/15/6/2100},
ISSN = {1099-4300},
DOI = {10.3390/e15062100}
}

@Article{Dann2020,
AUTHOR = {Dann, Roie and Kosloff, Ronnie and Salamon, Peter},
TITLE = {Quantum Finite-Time Thermodynamics: Insight from a Single Qubit Engine},
JOURNAL = {Entropy},
VOLUME = {22},
YEAR = {2020},
NUMBER = {11},
ARTICLE-NUMBER = {1255},
URL = {https://www.mdpi.com/1099-4300/22/11/1255},
PubMedID = {33287023},
ISSN = {1099-4300},
DOI = {10.3390/e22111255}
}

@book{breuer2002theory,
  title={The theory of open quantum systems},
  author={Breuer, Heinz-Peter and Petruccione, Francesco},
  year={2002},
  publisher={Oxford University Press, New York},
  doi={10.1093/acprof:oso/9780199213900.001.0001}
}

@article{alicki1979,
doi = {10.1088/0305-4470/12/5/007},
url = {https://doi.org/10.1088/0305-4470/12/5/007},
year = {1979},
month = {may},
publisher = {},
volume = {12},
number = {5},
pages = {L103},
author = {R Alicki},
title = {The quantum open system as a model of the heat engine},
journal = {J. Phys. A: Math. Gen.}
}

@article{uzdin2015,
  title={Equivalence of quantum heat machines, and quantum-thermodynamic signatures},
  author={Uzdin, Raam and Levy, Amikam and Kosloff, Ronnie},
  journal={Phys. Rev. X},
  volume={5},
  number={3},
  pages={031044},
  year={2015},
  publisher={APS},
  doi={10.1103/PhysRevX.5.031044},
  url={https://link.aps.org/doi/10.1103/PhysRevX.5.031044}
}

@article{PhysRevLett.113.260601,
  title = {Irreversible Work and Inner Friction in Quantum Thermodynamic Processes},
  author = {Plastina, F. and Alecce, A. and Apollaro, T. J. G. and Falcone, G. and Francica, G. and Galve, F. and Lo Gullo, N. and Zambrini, R.},
  journal = {Phys. Rev. Lett.},
  volume = {113},
  issue = {26},
  pages = {260601},
  numpages = {5},
  year = {2014},
  month = {Dec},
  publisher = {American Physical Society},
  doi = {10.1103/PhysRevLett.113.260601},
  url = {https://link.aps.org/doi/10.1103/PhysRevLett.113.260601}
}

@article{PhysRevLett.125.180603,
  title = {Quantum Coherence and Ergotropy},
  author = {Francica, G. and Binder, F. C. and Guarnieri, G. and Mitchison, M. T. and Goold, J. and Plastina, F.},
  journal = {Phys. Rev. Lett.},
  volume = {125},
  issue = {18},
  pages = {180603},
  numpages = {8},
  year = {2020},
  month = {Oct},
  publisher = {American Physical Society},
  doi = {10.1103/PhysRevLett.125.180603},
  url = {https://link.aps.org/doi/10.1103/PhysRevLett.125.180603}
}

@article{RevModPhys.83.771,
  title = {Colloquium: Quantum fluctuation relations: Foundations and applications},
  author = {Campisi, Michele and H\"anggi, Peter and Talkner, Peter},
  journal = {Rev. Mod. Phys.},
  volume = {83},
  issue = {3},
  pages = {771--791},
  numpages = {0},
  year = {2011},
  month = {Jul},
  publisher = {American Physical Society},
  doi = {10.1103/RevModPhys.83.771},
  url = {https://link.aps.org/doi/10.1103/RevModPhys.83.771}
}

@article{RevModPhys.81.1665,
  title = {Nonequilibrium fluctuations, fluctuation theorems, and counting statistics in quantum systems},
  author = {Esposito, Massimiliano and Harbola, Upendra and Mukamel, Shaul},
  journal = {Rev. Mod. Phys.},
  volume = {81},
  issue = {4},
  pages = {1665--1702},
  numpages = {0},
  year = {2009},
  month = {Dec},
  publisher = {American Physical Society},
  doi = {10.1103/RevModPhys.81.1665},
  url = {https://link.aps.org/doi/10.1103/RevModPhys.81.1665}
}

@incollection{takens1981,
  title={Detecting strange attractors in turbulence},
  author={Takens, Floris},
  booktitle={Dynamical systems and turbulence, Warwick 1980},
  pages={366--381},
  year={1981},
  publisher={Springer},
  doi={10.1007/BFb0091924}
}

@article{edelsbrunner2008,
  title={Persistent homology---a survey},
  author={Edelsbrunner, Herbert and Harer, John},
  journal={Contemp. Math.},
  volume={453},
  pages={257--282},
  year={2008},
  publisher={Providence, RI: American Mathematical Society},
  doi={10.1090/conm/453/08802}
}

@article{carlsson2009,
  title={Topology and data},
  author={Carlsson, Gunnar},
  journal={Bull. Amer. Math. Soc.},
  volume={46},
  number={2},
  pages={255--308},
  year={2009},
  doi={10.1090/S0273-0979-09-01249-X}
}

@article{macpherson2020,
  title={Topological data analysis of topological quantum phases},
  author={Sale, N and Giansiracusa, J and MacPherson, B},
  journal={Phys. Rev. Res.},
  volume={2},
  number={1},
  pages={013005},
  year={2020},
  publisher={APS},
  doi={10.1103/PhysRevResearch.2.013005},
  url={https://link.aps.org/doi/10.1103/PhysRevResearch.2.013005}
}

@article{Vinjanampathy01102016,
author = {Sai Vinjanampathy and Janet Anders},
title = {Quantum thermodynamics},
journal = {Contemp. Phys.},
volume = {57},
number = {4},
pages = {545--579},
year = {2016},
publisher = {Taylor \& Francis},
doi = {10.1080/00107514.2016.1201896},
URL = { https://doi.org/10.1080/00107514.2016.1201896}
}

@article{Goold_2016,
doi = {10.1088/1751-8113/49/14/143001},
url = {https://doi.org/10.1088/1751-8113/49/14/143001},
year = {2016},
month = {feb},
publisher = {IOP Publishing},
volume = {49},
number = {14},
pages = {143001},
author = {Goold, John and Huber, Marcus and Riera, Arnau and Rio, Lídia del and Skrzypczyk, Paul},
title = {The role of quantum information in thermodynamics—a topical review},
journal = {J. Phys. A: Math. Theor.}
}

@article{PhysRevLett.123.240601,
  title = {Experimental Characterization of a Spin Quantum Heat Engine},
  author = {Peterson, John P. S. and Batalh\~ao, Tiago B. and Herrera, Marcela and Souza, Alexandre M. and Sarthour, Roberto S. and Oliveira, Ivan S. and Serra, Roberto M.},
  journal = {Phys. Rev. Lett.},
  volume = {123},
  issue = {24},
  pages = {240601},
  numpages = {7},
  year = {2019},
  month = {Dec},
  publisher = {American Physical Society},
  doi = {10.1103/PhysRevLett.123.240601},
  url = {https://link.aps.org/doi/10.1103/PhysRevLett.123.240601}
}

@article{Johannes2016,
author = {Johannes Roßnagel  and Samuel T. Dawkins  and Karl N. Tolazzi  and Obinna Abah  and Eric Lutz  and Ferdinand Schmidt-Kaler  and Kilian Singer },
title = {A single-atom heat engine},
journal = {Science},
volume = {352},
number = {6283},
pages = {325-329},
year = {2016},
doi = {10.1126/science.aad6320},
URL = {https://www.science.org/doi/abs/10.1126/science.aad6320}
}

@article{PhysRevLett.122.110601,
  title = {Experimental Demonstration of Quantum Effects in the Operation of Microscopic Heat Engines},
  author = {Klatzow, James and Becker, Jonas N. and Ledingham, Patrick M. and Weinzetl, Christian and Kaczmarek, Krzysztof T. and Saunders, Dylan J. and Nunn, Joshua and Walmsley, Ian A. and Uzdin, Raam and Poem, Eilon},
  journal = {Phys. Rev. Lett.},
  volume = {122},
  issue = {11},
  pages = {110601},
  numpages = {6},
  year = {2019},
  month = {Mar},
  publisher = {American Physical Society},
  doi = {10.1103/PhysRevLett.122.110601},
  url = {https://link.aps.org/doi/10.1103/PhysRevLett.122.110601}
}

@article{PhysRevLett.127.200602,
  title = {Periodically Driven Quantum Thermal Machines from Warming up to Limit Cycle},
  author = {Liu, Junjie and Jung, Kenneth A. and Segal, Dvira},
  journal = {Phys. Rev. Lett.},
  volume = {127},
  issue = {20},
  pages = {200602},
  numpages = {7},
  year = {2021},
  month = {Nov},
  publisher = {American Physical Society},
  doi = {10.1103/PhysRevLett.127.200602},
  url = {https://link.aps.org/doi/10.1103/PhysRevLett.127.200602}
}

@article{PhysRevE.76.031105,
  title = {Quantum thermodynamic cycles and quantum heat engines},
  author = {Quan, H. T. and Liu, Yu-xi and Sun, C. P. and Nori, Franco},
  journal = {Phys. Rev. E},
  volume = {76},
  issue = {3},
  pages = {031105},
  numpages = {18},
  year = {2007},
  month = {Sep},
  publisher = {American Physical Society},
  doi = {10.1103/PhysRevE.76.031105},
  url = {https://link.aps.org/doi/10.1103/PhysRevE.76.031105}
}

@article{PhysRevB.96.104304,
  title = {Quantum efficiency bound for continuous heat engines coupled to noncanonical reservoirs},
  author = {Agarwalla, Bijay Kumar and Jiang, Jian-Hua and Segal, Dvira},
  journal = {Phys. Rev. B},
  volume = {96},
  issue = {10},
  pages = {104304},
  numpages = {6},
  year = {2017},
  month = {Sep},
  publisher = {American Physical Society},
  doi = {10.1103/PhysRevB.96.104304},
  url = {https://link.aps.org/doi/10.1103/PhysRevB.96.104304}
}

@article{PhysRevLett.2.262,
  title = {Three-Level Masers as Heat Engines},
  author = {Scovil, H. E. D. and Schulz-DuBois, E. O.},
  journal = {Phys. Rev. Lett.},
  volume = {2},
  issue = {6},
  pages = {262--263},
  numpages = {0},
  year = {1959},
  month = {Mar},
  publisher = {American Physical Society},
  doi = {10.1103/PhysRevLett.2.262},
  url = {https://link.aps.org/doi/10.1103/PhysRevLett.2.262}
}

@article{PhysRevLett.129.100603,
  title = {Experimental Realization of a Quantum Refrigerator Driven by Indefinite Causal Orders},
  author = {Nie, Xinfang and Zhu, Xuanran and Huang, Keyi and Tang, Kai and Long, Xinyue and Lin, Zidong and Tian, Yu and Qiu, Chudan and Xi, Cheng and Yang, Xiaodong and Li, Jun and Dong, Ying and Xin, Tao and Lu, Dawei},
  journal = {Phys. Rev. Lett.},
  volume = {129},
  issue = {10},
  pages = {100603},
  numpages = {6},
  year = {2022},
  month = {Sep},
  publisher = {American Physical Society},
  doi = {10.1103/PhysRevLett.129.100603},
  url = {https://link.aps.org/doi/10.1103/PhysRevLett.129.100603}
}

@article{PhysRevE.68.016101,
  title = {Quantum four-stroke heat engine: Thermodynamic observables in a model with intrinsic friction},
  author = {Feldmann, Tova and Kosloff, Ronnie},
  journal = {Phys. Rev. E},
  volume = {68},
  issue = {1},
  pages = {016101},
  numpages = {18},
  year = {2003},
  month = {Jul},
  publisher = {American Physical Society},
  doi = {10.1103/PhysRevE.68.016101},
  url = {https://link.aps.org/doi/10.1103/PhysRevE.68.016101}
}

@article{kato1950,
author = {Kato ,Tosio},
title = {On the Adiabatic Theorem of Quantum Mechanics},
journal = {J. Phys. Soc. Jpn.},
volume = {5},
number = {6},
pages = {435-439},
year = {1950},
doi = {10.1143/JPSJ.5.435},
URL = { https://doi.org/10.1143/JPSJ.5.435}
}

@book{messiah2014quantum,
  title={Quantum mechanics},
  author={Messiah, Albert},
  year={2014},
  publisher={Courier Corporation}
}

@article{Rezek_2006,
doi = {10.1088/1367-2630/8/5/083},
url = {https://doi.org/10.1088/1367-2630/8/5/083},
year = {2006},
month = {may},
publisher = {},
volume = {8},
number = {5},
pages = {83},
author = {Rezek, Yair and Kosloff, Ronnie},
title = {Irreversible performance of a quantum harmonic heat engine},
journal = {New J. Phys.}
}

@article{PhysRevLett.105.150603,
  title = {Efficiency at Maximum Power of Low-Dissipation Carnot Engines},
  author = {Esposito, Massimiliano and Kawai, Ryoichi and Lindenberg, Katja and Van den Broeck, Christian},
  journal = {Phys. Rev. Lett.},
  volume = {105},
  issue = {15},
  pages = {150603},
  numpages = {4},
  year = {2010},
  month = {Oct},
  publisher = {American Physical Society},
  doi = {10.1103/PhysRevLett.105.150603},
  url = {https://link.aps.org/doi/10.1103/PhysRevLett.105.150603}
}

@article{PhysRevLett.109.203006,
  title = {Single-Ion Heat Engine at Maximum Power},
  author = {Abah, O. and Ro\ss{}nagel, J. and Jacob, G. and Deffner, S. and Schmidt-Kaler, F. and Singer, K. and Lutz, E.},
  journal = {Phys. Rev. Lett.},
  volume = {109},
  issue = {20},
  pages = {203006},
  numpages = {6},
  year = {2012},
  month = {Nov},
  publisher = {American Physical Society},
  doi = {10.1103/PhysRevLett.109.203006},
  url = {https://link.aps.org/doi/10.1103/PhysRevLett.109.203006}
}

@misc{dechecchi2025dynamics,
      title={Quantum trajectories and reduced dynamics in time-correlated environments}, 
      author={Pietro De Checchi and Federico Gallina and Barbara Fresch and Giulio G. Giusteri},
      eprint={2507.17864},
      archivePrefix={arXiv},
    year={2026},
      url={https://arxiv.org/abs/2507.17864}, 
}

@misc{cantone2025machine,
      title={Machine Learning-aided Optimal Control of a noisy qubit}, 
      author={Riccardo Cantone and Shreyasi Mukherjee and Luigi Giannelli and Elisabetta Paladino and Giuseppe Falci},
      year={2025},
      eprint={2507.14085},
      archivePrefix={arXiv},
      primaryClass={quant-ph},
      url={https://arxiv.org/abs/2507.14085}, 
}

@article{aguilar2008effect,
  title={The effect of classical noise on a quantum two-level system},
  author={Aguilar, Jean-Philippe and Berglund, Nils},
  journal={J. Math. Phys.},
  volume={49},
  number={10},
  pages={102102},
  year={2008},
  publisher={AIP Publishing},
  doi={10.1063/1.2993981},
  url={https://doi.org/10.1063/1.2988180}
}

@article{stefanatos2020robustness,
  title={Robustness of STIRAP Shortcuts under Ornstein-Uhlenbeck Noise in the Energy Levels},
  author={Stefanatos, Dionisis and Blekos, Kostas and Paspalakis, Emmanuel},
  journal={Appl. Sci.},
  volume={10},
  number={5},
  pages={1580},
  year={2020},
  publisher={MDPI},
  doi={10.3390/app10051580},
  url={https://doi.org/10.3390/app10051580}
}

@article{adams2017,
  title={Persistence images: A stable vector representation of persistent homology},
  author={Adams, Henry and Emerson, Tegan and Kirby, Michael and Neville, Rachel and Peterson, Chris and Shipman, Patrick and Chepushtanova, Sofya and Hanson, Eric and Motta, Francis and Ziegelmeier, Lori},
  journal={J. Mach. Learn. Res.},
  volume={18},
  number={8},
  pages={1--35},
  year={2017},
  publisher={JMLR.org},
  doi={10.48550/arXiv.1507.06217},
  url={http://jmlr.org/papers/v18/16-337.html}
}

@inproceedings{chazal2014,
  title={Stochastic convergence of persistence landscapes and silhouettes},
  author={Chazal, Fr{\'e}d{\'e}ric and Fasy, Brittany Terese and Lecci, Fabrizio and Rinaldo, Alessandro and Wasserman, Larry},
  booktitle={Proceedings of the Thirtieth Annual Symposium on Computational Geometry},
  pages={474--483},
  year={2014},
  publisher={Association for Computing Machinery (ACM)},
  doi={10.1145/2582112.2582128},
  url={https://doi.org/10.1145/2582112.2582128}
}

@article{blais2021,
  title={Circuit quantum electrodynamics},
  author={Blais, Alexandre and Grimsmo, Arne L and Girvin, Steven M and Wallraff, Andreas},
  journal={Rev. Mod. Phys.},
  volume={93},
  number={2},
  pages={025005},
  year={2021},
  publisher={APS},
  doi={10.1103/RevModPhys.93.025005},
  url={https://link.aps.org/doi/10.1103/RevModPhys.93.025005}
}

@article{bruzewicz2019,
  title={Trapped-ion quantum computing: Progress and challenges},
  author={Bruzewicz, Colin D and Chiaverini, John and McConnell, Robert and Sage, Jeremy M},
  journal={Appl. Phys. Rev. },
  volume={6},
  number={2},
  pages={021314},
  year={2019},
  publisher={AIP Publishing},
  doi={10.1063/1.5088164},
  url={https://doi.org/10.1063/1.5088164}
}

@article{Cao2023,
  title = {Unravelling quantum chaos using persistent homology},
  author = {Cao, Harvey and Leykam, Daniel and Angelakis, Dimitris G.},
  journal = {Phys. Rev. E},
  volume = {107},
  issue = {4},
  pages = {044204},
  numpages = {10},
  year = {2023},
  month = {Apr},
  publisher = {American Physical Society},
  doi = {10.1103/PhysRevE.107.044204},
  url = {https://link.aps.org/doi/10.1103/PhysRevE.107.044204}
}

@article{Spitz2021,
  title = {Finding self-similar behavior in quantum many-body dynamics via persistent homology},
  author = {Spitz, Daniel and Berges, J\"{u}rgen and Oberthaler, Markus and Wienhard, Anna},
  journal = {SciPost Phys.},
  volume = {11},
  issue = {3},
  pages = {060},
  year = {2021},
  month = {Sep},
  publisher = {SciPost},
  doi = {10.21468/SciPostPhys.11.3.060},
  url = {https://scipost.org/10.21468/SciPostPhys.11.3.060}
}

@article{Leykam2023,
  title = {Topological data analysis and machine learning},
  author = {Leykam, Daniel and Angelakis, Dimitris G.},
  journal = {Adv. Phys.: X},
  volume = {8},
  number = {1},
  pages = {2202331},
  year = {2023},
  publisher = {Taylor & Francis},
  doi = {10.1080/23746149.2023.2202331},
  url = {https://doi.org/10.1080/23746149.2023.2202331}
}

@article{Scali2024,
  title = {Quantum topological data analysis via the estimation of the density of states},
  author = {Scali, Stefano and Umeano, Chukwudubem and Kyriienko, Oleksandr},
  journal = {Phys. Rev. A},
  volume = {110},
  issue = {4},
  pages = {042616},
  numpages = {26},
  year = {2024},
  month = {Oct},
  publisher = {American Physical Society},
  doi = {10.1103/PhysRevA.110.042616},
  url = {https://link.aps.org/doi/10.1103/PhysRevA.110.042616}
}

@article{gvys-hl8h,
  title = {Provable Quantum Speedups for Computing Persistence in Topological Data Analysis},
  author = {Gyurik, Casper and Schmidhuber, Alexander and King, Robbie and Dunjko, Vedran and Hayakawa, Ryu},
  journal = {PRX Quantum},
  volume = {7},
  issue = {2},
  pages = {020361},
  numpages = {18},
  year = {2026},
  month = {Jun},
  publisher = {American Physical Society},
  doi = {10.1103/gvys-hl8h},
  url = {https://link.aps.org/doi/10.1103/gvys-hl8h}
}

@article{Spitz2025,
  title = {Topological data analysis of the deconfinement transition in SU(3) lattice gauge theory},
  author = {Spitz, Daniel and Urban, Julian M. and Pawlowski, Jan M.},
  journal = {Phys. Rev. D},
  volume = {111},
  issue = {11},
  pages = {114519},
  numpages = {23},
  year = {2025},
  month = {Jun},
  publisher = {American Physical Society},
  doi = {10.1103/PhysRevD.111.114519},
  url = {https://link.aps.org/doi/10.1103/PhysRevD.111.114519}
}

@article{berry2009transitionless,
  author  = {Berry, M. V.},
  title   = {Transitionless quantum driving},
  journal = {J. Phys. A: Math. Theo.},
  volume  = {42},
  pages   = {365303},
  year    = {2009},
  doi     = {10.1088/1751-8113/42/36/365303}
}

@incollection{torrontegui2013shortcuts,
  author    = {Torrontegui, E. and Ibanez, S. and Martinez-Garaot, S. and Modugno, M. and del Campo, A. and Guery-Odelin, D. and Ruschhaupt, A. and Chen, X. and Muga, J. G.},
  title     = {Shortcuts to Adiabaticity},
  booktitle = {Advances in Atomic, Molecular, and Optical Physics},
  volume    = {62},
  pages     = {117--169},
  publisher = {Academic Press},
  year      = {2013},
  doi       = {10.1016/B978-0-12-408090-4.00002-5}
}

@article{cakmak2019spin,
  author  = {{\c{C}}akmak, Bar{\i}\c{s} and M{\"u}stecapl{\i}o{\u{g}}lu, {\"O}zg{\"u}r E.},
  title   = {Spin quantum heat engines with shortcuts to adiabaticity},
  journal = {Physical Review E},
  volume  = {99},
  pages   = {032108},
  year    = {2019},
  doi     = {10.1103/PhysRevE.99.032108}
}

@article{delcampo2014superadiabatic,
  author  = {del Campo, Adolfo and Goold, John and Paternostro, Mauro},
  title   = {More bang for your buck: Super-adiabatic quantum engines},
  journal = {Scientific Reports},
  volume  = {4},
  pages   = {6208},
  year    = {2014},
  doi     = {10.1038/srep06208}
}

@article{pedram2023quantum,
  author  = {Pedram, Ali and Kad{\i}o{\u{g}}lu, S. C. and Kabak{\c{c}}{\i}o{\u{g}}lu, Alkan and M{\"u}stecapl{\i}o{\u{g}}lu, {\"O}zg{\"u}r E.},
  title   = {A quantum Otto engine with shortcuts to thermalization and adiabaticity},
  journal = {New Journal of Physics},
  volume  = {25},
  pages   = {113014},
  year    = {2023},
  doi     = {10.1088/1367-2630/ad08f6}
}

@misc{Maden2026TEMSupplement,
  author       = {Mira{\c{c}} Kerem Maden and Asghar Ullah and Baris Coskunuzer and {\"O}zg{\"u}r E. M{\"u}stecapl{\i}o{\u{g}}lu},
  title        = {Supplementary Material for Topological Engine Monitor},
  year         = {2026},
  howpublished = {\url{https://github.com/mk-maden/topological-engine-monitor}},
  note         = {GitHub repository},
}

\end{document}